\documentclass{aa}

\newcommand{\AK}[1]{{\color{black}#1}}

\newif\iffigure
\figurefalse
\figuretrue

\usepackage{graphicx,natbib,url,twoopt}
\usepackage[varg]{txfonts}           
\usepackage{hyperref}              
\usepackage{pdfcomment}              
\usepackage{acronym}                 
\usepackage{comment}
\usepackage{cases}
\usepackage{empheq}                  
\usepackage{here}
\usepackage{bm}
\usepackage[version=3]{mhchem}
\hypersetup{
 colorlinks=true,%
 linkcolor=blue,
 citecolor=blue,
}

\usepackage{lineno}
\newcommand*\patchAmsMathEnvironmentForLineno[1]{
  \expandafter\let\csname old#1\expandafter\endcsname\csname #1\endcsname
  \expandafter\let\csname oldend#1\expandafter\endcsname\csname end#1\endcsname
  \renewenvironment{#1}
     {\linenomath\csname old#1\endcsname}
     {\csname oldend#1\endcsname\endlinenomath}}
\newcommand*\patchBothAmsMathEnvironmentsForLineno[1]{
  \patchAmsMathEnvironmentForLineno{#1}
  \patchAmsMathEnvironmentForLineno{#1*}}
\AtBeginDocument{
\patchBothAmsMathEnvironmentsForLineno{equation}
\patchBothAmsMathEnvironmentsForLineno{align}
\patchBothAmsMathEnvironmentsForLineno{flalign}
\patchBothAmsMathEnvironmentsForLineno{alignat}
\patchBothAmsMathEnvironmentsForLineno{gather}
\patchBothAmsMathEnvironmentsForLineno{multline}
}

\bibpunct{(}{)}{;}{a}{}{,}    

\makeatletter
\newcommand{\bibnote}[2]{\global\@namedef{#1note}{#2}}
\newcommand{\biblink}[2]{\global\@namedef{#1link}{#2}}
\newcommand{\Tabref}[1]{Table~\ref{#1}}
\newcommand{\Equref}[1]{Eq.~(\ref{#1})}
\newcommand{\Figref}[1]{Fig.~\ref{#1}}
\makeatother

\makeatletter
 \newcommandtwoopt{\citeads}[3][][]{%
   \nonstopmode
   \href{http://adsabs.harvard.edu/abs/#3}%
        {\def\hyper@linkstart##1##2{}%
         \let\hyper@linkend\@empty\citealp[#1][#2]{#3}}
   \biblink{#3}{\href{http://adsabs.harvard.edu/abs/#3}{ADS}}%
   \errorstopmode}            
 \newcommandtwoopt{\citepads}[3][][]{%
   \nonstopmode
   \href{http://adsabs.harvard.edu/abs/#3}%
        {\def\hyper@linkstart##1##2{}%
         \let\hyper@linkend\@empty\citep[#1][#2]{#3}}
   \biblink{#3}{\href{http://adsabs.harvard.edu/abs/#3}{ADS}}%
   \errorstopmode}            
 \newcommandtwoopt{\citetads}[3][][]{%
   \nonstopmode
   \href{http://adsabs.harvard.edu/abs/#3}%
        {\def\hyper@linkstart##1##2{}%
         \let\hyper@linkend\@empty\citet[#1][#2]{#3}}
   \biblink{#3}{\href{http://adsabs.harvard.edu/abs/#3}{ADS}}%
   \errorstopmode}            
 \newcommandtwoopt{\citeyearads}[3][][]{%
   \nonstopmode
   \href{http://adsabs.harvard.edu/abs/#3}%
        {\def\hyper@linkstart##1##2{}%
         \let\hyper@linkend\@empty\citeyear[#1][#2]{#3}}
   \biblink{#3}{\href{http://adsabs.harvard.edu/abs/#3}{ADS}}%
   \errorstopmode}            
\makeatother

\newacro{ADS}{Astrophysics Data System}
\newacro{NLTE}{non-local thermodynamic equilibrium}
\newacro{NASA}{National Aeronautics and Space Administration}

\defcitealias{Kuwahara:2020}{Paper {\rm I}}

\begin{document}
\authorrunning{A. Kuwahara and H. Kurokawa}
\titlerunning{Influences of protoplanet-induced three-dimensional gas flow on pebble accretion. $\rm\,I\hspace{-.1em}I$.}
   \title{Influences of protoplanet-induced three-dimensional gas \\ 
   flow on pebble accretion \\ $\rm\,I\hspace{-.1em}I\,$. Headwind regime}
   \subtitle{}

          \author{Ayumu Kuwahara\inst{1,2}
          \thanks{\email{kuwahara@elsi.jp}} 
          \and Hiroyuki Kurokawa\inst{2}}

   \institute{Department of Earth and Planetary Sciences, Tokyo Institute of Technology, Ookayama, Meguro-ku, Tokyo, 152-8551, Japan
         \and
             Earth-Life Science Institute, Tokyo Institute of Technology, Ookayama, Meguro-ku, Tokyo, 152-8550, Japan}

   \date{Received September XXX; accepted YYY}

 
  \abstract
   {Pebble accretion is one of the major theories in planet formation. Aerodynamically small particles, called pebbles, are highly affected by the gas flow. A growing planet embedded in a protoplanetary disk induces three-dimensional (3D) gas flow. In our previous study, Paper I, we focused on the shear regime of pebble accretion, and investigated the influence of planet-induced gas flow on pebble accretion. In Paper I, we found that pebble accretion is inefficient in the planet-induced gas flow compared to that in the unperturbed flow, in particular when ${\rm St}\lesssim10^{-3}$, where St is the Stokes number. }
   {Following Paper I, we investigate the influence of planet-induced gas flow on pebble accretion. We consider the headwind of the gas, which is not included in Paper I. We extend our study to the headwind regime of pebble accretion in this study.}
   {Assuming a nonisothermal, inviscid sub-Keplerian gas disk, we perform 3D hydrodynamical simulations on the spherical polar grid which has a planet with the dimensionless mass, $m=R_{\rm Bondi}/H$, located at its center, where $R_{\rm Bondi}$ and $H$ are the Bondi radius and the disk scale height. We then numerically integrate the equation of motion of pebbles in 3D using hydrodynamical simulation data.}
   {We first divide the planet-induced gas flow into two regimes: the flow shear and flow headwind regimes. In the flow shear regime, where the planet-induced gas flow has a vertically rotational symmetric structure, we find that the outcome is identical to that obtained in Paper I. In the flow headwind regime, the strong headwind of the gas breaks the symmetric structure of the planet-induced gas flow. In the flow headwind regime, we find that the trajectories of pebbles with ${\rm St}\lesssim10^{-3}$ in the planet-induced gas flow differ significantly from those in the unperturbed flow. The recycling flow, where gas from the disk enters the gravitational sphere at low latitudes and exits at high latitudes, gathers pebbles around the planet. We derive the flow transition mass analytically, $m_{\rm t,\,flow}$, which discriminates between the flow headwind and flow shear regimes. From the relation between $m,\,m_{\rm t,\,flow}$ and $m_{\rm t,\,peb}$, where $m_{\rm t,\,peb}$ is the transition mass of the accretion regime of pebbles, we classify the results obtained in both Paper I and this study into four groups. In particular, only when the Stokes gas drag law is adopted and $m<m_{\rm t,\,flow}<m_{\rm t,\,peb}$, where the accretion and flow regime are both in the headwind regime, the accretion probability of pebbles with ${\rm St}\lesssim10^{-3}$ is enhanced in the planet-induced gas flow compared to that in the unperturbed flow.}
   {Combining our results with the spacial variety of turbulence strength and pebble size in a disk, we conclude that the planet-induced gas flow \AK{still allows for} pebble accretion in the early stage of planet formation. Suppression of pebble accretion due to the planet-induced gas flow occurs only in the late stage of planet formation, in particular in the inner region of the disk. This may be helpful to explain the distribution of exoplanets and the architecture of the Solar System, both of which have small inner and large outer planets.}
   

   \maketitle
%

\section{Introduction}\label{sec:Introduction}
Recent hydrodynamical simulations have revealed that a planet embedded in a protoplanetary disk induces gas flow with a complex 3D structure \cite[]{Ormel:2015b,Fung:2015,Lambrechts:2017,Cimerman:2017,Kurokawa:2018,Kuwahara:2019,Bethune:2019,Fung:2019}. The anterior-posterior horseshoe flows extending in the orbital direction of the planet have a characteristic vertical structure like a column. A substantial amount of gas from the disk enters the gravitational sphere of the planet (inflow), and exits it (outflow), causing atmospheric recycling. Qualitatively, the 3D flow structure depends on the magnitude of the deviation of the speed of the gas from Keplerian rotation. In a Keplerian disk, the 3D planet-induced flow has a vertically rotational symmetric structure, but the symmetry is broken in a sub-Keplerian disk \cite[]{Ormel:2015b,Kurokawa:2018}. The induced gas flow affects pebble accretion and may alter the accretion probability of pebbles. It has been recognized that the accretion rate of small particles ($\sim100$ $\mu$m--$1$ mm) is reduced in the 2D planet-induced flow \cite[]{Ormel:2013}. Pebble accretion in the 3D planet-induced gas flow becomes more complicated. \cite{Popovas:2018a,Popovas:2018b} incorporated pebbles into their hydrodynamical simulations and found that small particles ($10\,\mu$m--$1$ cm) move away from the planet in the horseshoe flow and avoid accretion onto Earth- and Mars-sized planets.

Assuming a Keplerian disk, \cite{Kuwahara:2020} (hereafter \citetalias{Kuwahara:2020}) performed orbital calculation of pebbles in 3D using hydrodynamical simulation data, finding that the 3D planet-induced gas flow affects pebble accretion significantly. In \citetalias{Kuwahara:2020}, planets of between three Mars masses and three Earth masses orbiting a solar-mass star at 1 au, $\sim0.3$--$3\,M_{\oplus}$, are considered. The contribution of the headwind of the gas was \AK{not investigated}. The shear regime of pebble accretion\footnote{In this study, we used "headwind" and "shear" regimes as the names to distinguish the pebble accretion regimes, which are used in \cite[]{Ormel:2017-pebble}. These regimes are referred to as "Bondi" and "Hill" regimes in \cite{Lambrechts:2012}.} was only considered in \citetalias{Kuwahara:2020}, where the accretion radius for pebble accretion can be characterized by the size of the Hill radius, and the approach velocity of pebbles is set by the shear velocity \cite[]{Lambrechts:2012,Ormel:2017-pebble,Johansen:2017}. When pebbles are aerodynamically small, those coming from within the vicinity of the planetary orbit move away from the planet along the horseshoe flows. The outflow of the gas at the midplane region deflects the pebble trajectories and inhibits small pebbles from accreting. The pebbles coming from a window between the horseshoe and the shear regions can accrete onto the planet. Thus, the width of the accretion window in the planet-induced gas flow is narrower than that in the unperturbed flow. 

The accretion probability of pebbles, which is an important parameter to control the outcome of the pebble-driven planet formation model, is affected by the planet-induced gas flow. For a planet with $\sim0.3\,M_{\oplus}$, the accretion probability in the planet-induced gas flow is smaller than that in the unperturbed flow \AK{\citepalias{Kuwahara:2020}}. This is caused by the reduction of the width of the accretion window. When the planetary mass is larger than $0.3\,M_{\oplus}$, the accretion probability in the planet-induced gas flow is comparable to that in the unperturbed flow, except for when the pebbles are well coupled to the gas. As the planetary mass increases, the width of the horseshoe region increases. Pebbles with high relative velocity accrete onto the planet. Thus, the reduction of the width of the accretion window and the increase of relative velocity cancel each other out. 

In the protoplanetary disks, the disk gas rotates slower than Keplerian velocity due to the existence of the global pressure gradient. In \citetalias{Kuwahara:2020}, we focused on large planetary masses, $\gtrsim0.3\,M_{\oplus}$, for which the influence of the headwind on pebble accretion is negligible. In an early phase of the planetary growth, however, pebble accretion proceeds in the headwind regime, where the accretion radius for pebble accretion can be characterized by the size of the Bondi radius, and the approach velocity of pebbles is set by the sub-Keplerian speed \cite[]{Lambrechts:2012,Ormel:2017-pebble,Johansen:2017}. Furthermore, the influence of the 3D planet-induced gas flow whose vertically rotational symmetry is broken due to the strong headwind of the gas is still unclear. In Paper $\rm I\hspace{-.1em}I\,$, we extend our study in \citetalias{Kuwahara:2020} to the headwind regime.

The structure of this paper is as follows. In Sect. \ref{sec:methods} we describe the numerical method. In Sect. \ref{sec:result} we show the results obtained from a series of simulations. In Sect. \ref{sec:discussion} we discuss the implications for planet formation. We summarize in Sect. \ref{sec:conclusion}.
\section{Methods}\label{sec:methods}

\iffigure
 \begin{figure*}[htbp]
 \resizebox{\hsize}{!}
 {\includegraphics{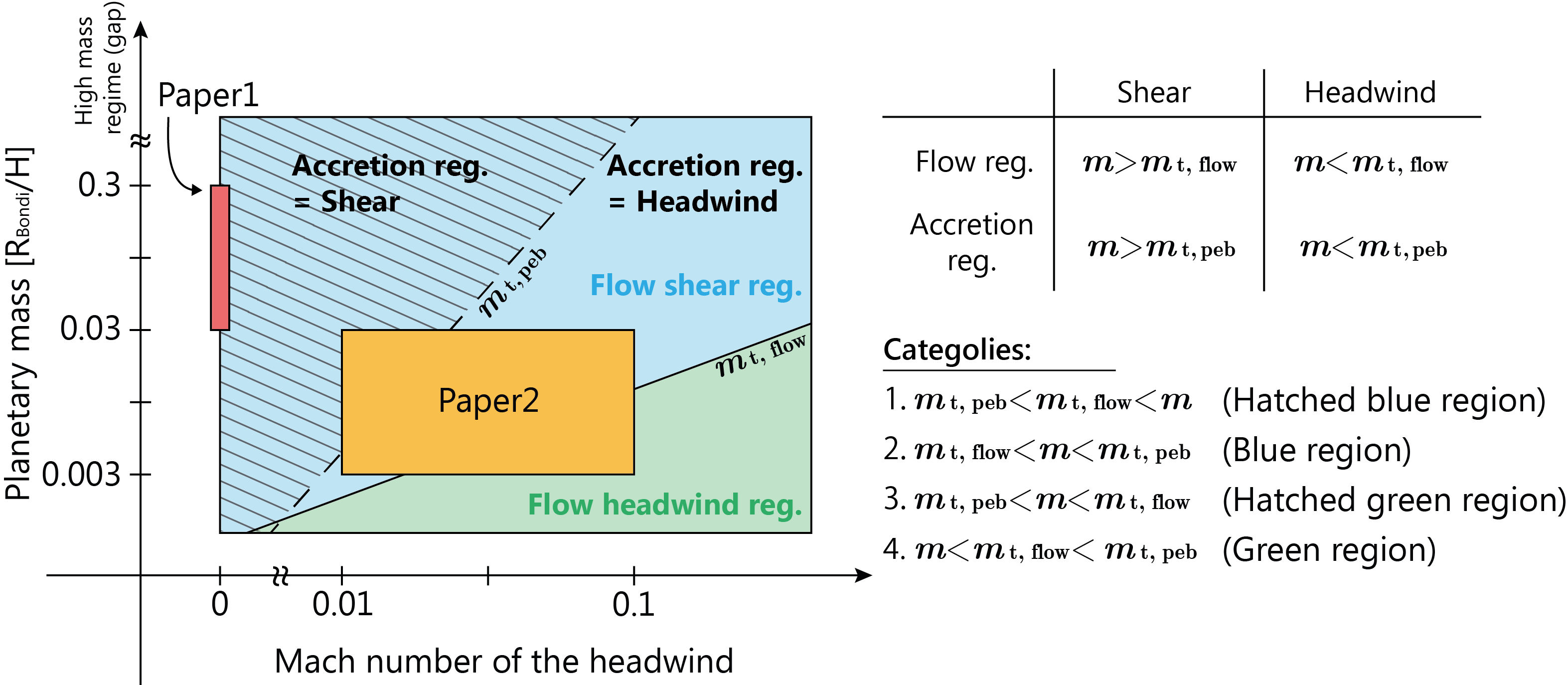}} 
 \caption{\AK{Summary of parameter surveys in \citetalias{Kuwahara:2020} and $\rm\,I\hspace{-.1em}I\,$. The parameter spaces investigated in both \citetalias{Kuwahara:2020} and $\rm\,I\hspace{-.1em}I\,$ are shown in red and yellow filled squares, respectively. The vertical and horizontal axes are the dimensionless planetary mass, $m$ (\Equref{eq:m}), and the Mach number of the headwind of the gas, $\mathcal{M}_{\rm hw}$ (\Equref{eq:Mach}), respectively. The black solid and dashed lines correspond to the flow transition mass, $m_{\rm t,\,flow}$ (\Equref{eq:m-flow-2}), and the pebble transition mass, $m_{\rm t,\,peb}$ (\Equref{eq:dim-less-transition}), for ${\rm St}=10^{-3}$, where ${\rm St}$ is the Stokes number of pebbles (\Equref{eq:St}). Our study does not handle the gap opening \AK{by} high-mass planets. The classification for the flow and accretion regimes is shown in the upper right corner. The four categories consisting of combinations of the flow and accretion regimes are shown in the bottom right corner (see Sect. \ref{sec:flow-transition} and \Figref{fig:schematic} for the detailed description).}}
\label{fig:summary}
\end{figure*}
\fi

\begin{table*}[htpb]
\caption{List of parameters and regimes. The left column gives the planetary mass, the Stokes number, the Mach number of the headwind, the regime of the planet-induced gas flow (Sect. \ref{sec:clasification}), the regime of pebble accretion (Sect. \ref{sec:appendix1}), and the regime of the gas drag law (Sect. \ref{sec:orbital}), respectively. The middle and right columns show the range of the parameters and the flow, accretion, and the gas drag regimes investigated in both \citetalias{Kuwahara:2020} and $\rm\,I\hspace{-.1em}I\,$, respectively.}
\centering
\begin{tabular}{llll}\hline\hline
Parameter, Regime & \citetalias{Kuwahara:2020} & Paper $\rm\,I\hspace{-.1em}I$\\ \hline
$m$ (\Equref{eq:m}) & $0.03,\,0.1,\,0.3$ & $0.003,\,0.01,\,0.03$ \\
${\rm St}$ (\Equref{eq:St}) & $10^{-3}$--$10^0$ & $10^{-4}$--$10^{0}$\\
$\mathcal{M}_{\rm hw}$ (\Equref{eq:Mach}) & $0$ (Keplerian disk) & $0.01,\,0.03,\,0.1$ (sub-Keplerian disk)\\
Flow regime (\Equref{eq:m-flow-2}) & Shear & Shear \& Headwind \\
Accretion regime (Eqs. (\ref{eq:b-hw}) and (\ref{eq:b-sh})) & Shear & Shear \& Headwind \\
Gas drag regime (Eqs. (\ref{eq:Epstein-regime}) and (\ref{eq:Stokes-regime})) & Stokes \& Epstein & Stokes \& Epstein \\ 
\hline
\end{tabular}
\label{tab:2}
\end{table*}

\begin{table*}[htbp]
\caption{List of hydrodynamical simulations. The following columns give the simulation name, the size of the Bondi radius of the planet, the size of the Hill radius of the planet, the size of the inner boundary, the size of the outer boundary, the length of the calculation time, the dimensionless thermal relaxation timescale $\beta$, the Mach number of the headwind, and the regime of the planet-induced gas flow (Sect. \ref{sec:clasification}) respectively.}
\centering
\begin{tabular}{lcccccccl}\hline\hline
Name & $R_{\rm Bondi}$ & $R_{\rm Hill}$ & $r_{\rm inn}$ & $r_{\rm out}$ & $t_{\rm end}$ & $\beta$ & $\mathcal{M}_{\rm hw}$ & flow regime\\ \hline
\texttt{m0003-hw001} & 0.003 & 0.1 & 4.33$\times10^{-4}$ & 0.05 & 10 & $9\times10^{-4}$ & $0.01$ & flow shear\\
\texttt{m0003-hw003} & 0.003 & 0.1 & 4.33$\times10^{-4}$ & 0.05 & 10 & $9\times10^{-4}$ & $0.03$ & flow headwind\\
\texttt{m0003-hw01} & 0.003 & 0.1 & 4.33$\times10^{-4}$ & 0.05 & 10 & $9\times10^{-4}$ & $0.1$ & flow headwind\\
\texttt{m001-hw001} & 0.01 & 0.15 & 6.46$\times10^{-4}$ & 0.5  & 50 & 0.01 & $0.01$ & flow shear\\
\texttt{m001-hw003} & 0.01 & 0.15 & 6.46$\times10^{-4}$ & 0.5  & 50 & 0.01 & $0.03$ & flow shear\\
\texttt{m001-hw01} & 0.01 & 0.15 & 6.46$\times10^{-4}$ & 0.5  & 50 & 0.01 & $0.1$ & flow headwind\\
\texttt{m003-hw001} & 0.03 & 0.22 & 9.32$\times10^{-4}$ & 0.5 & 50 & 0.09 & $0.01$ & flow shear \\
\texttt{m003-hw003} & 0.03 & 0.22 & 9.32$\times10^{-4}$ & 0.5 & 50 & 0.09 & $0.03$ & flow shear \\
\texttt{m003-hw01} & 0.03 & 0.22 & 9.32$\times10^{-4}$ & 0.5 & 50 & 0.09 & $0.1$ & flow headwind \\\hline
\end{tabular}
\label{tab:1}
\end{table*}

\iffigure
 \begin{figure}[htbp]
 \resizebox{\hsize}{!}
 {\includegraphics{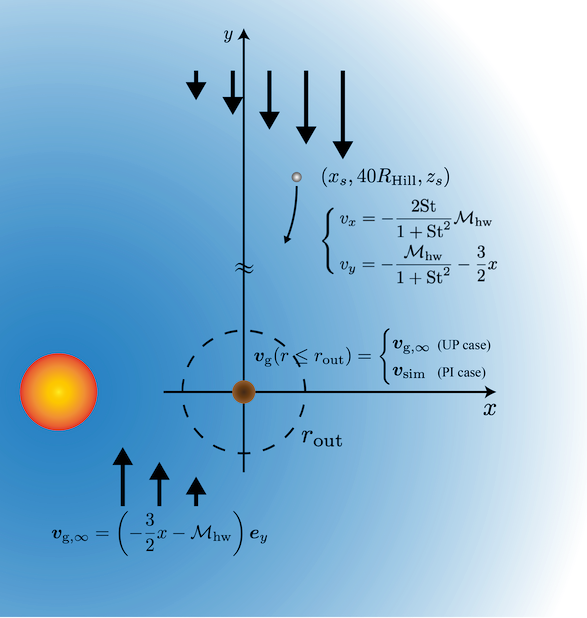}} 
 \caption{Schematic picture of orbital calculation of pebbles. A planet is located at the origin of the co-rotational frame. The dashed line represents the outer boundary of the hydrodynamical simulations. The starting point of orbital calculation is beyond the outer boundary of hydrodynamical simulations. Its $y$-component is fixed at $40R_{\rm Hill}$ \cite[]{Ida:1989}. The $x$- and $z$-coordinates of the starting point of pebbles, $x_{\rm s}$ and $z_{\rm s}$, are the parameters. The gas velocity is assumed to be the speed of the sub-Keplerian shear both inside and outside of $r_{\rm out}$ in shear flow case in the unperturbed flow (\texttt{UP-mXX-hwYY} case), but is switched to the gas velocity obtained from the hydrodynamical simulations within $r_{\rm out}$ in the planet-induced flow case in the Epstein regime (\texttt{PI-Epstein-mXX-hwYY} case), and in the Stokes regime (\texttt{PI-Stokes-mXX-hwYY} case).}
\label{fig:frame}
\end{figure}
\fi

\subsection{Model overview}
Most of our methods are the same as described in \citetalias{Kuwahara:2020}, except for the investigation of the headwind of the gas. In the following sections, we describe the differences from \citetalias{Kuwahara:2020} and emphasize the key points of our model. Through all of our simulations the length, times, velocities, and densities are normalized by the disk scale height $H$, the reciprocal of the orbital frequency $\Omega^{-1}$, the sound speed $c_{\rm s}$, and the unperturbed gas density at the location of the planet $\rho_{\rm disk}$, respectively. In this unit, the dimensionless planetary mass is given by
\begin{align}
m=\frac{R_{\rm Bondi}}{H}=\frac{GM_{\rm pl}}{c_{\rm s}^{3}/\Omega},\label{eq:m}
\end{align}
where $G$ is the gravitational constant, and $M_{\rm pl}$ is the mass of the planet. The Hill radius is given by $R_{\rm Hill}=(m/3)^{1/3}H$ in this unit. We assume the minimum mass solar nebula (MMSN) model \cite[]{Weidenschilling:1977,Hayashi:1985} when we convert the dimensionless quantities into dimensional ones.

\AK{We summarize the parameter spaces investigated in both \citetalias{Kuwahara:2020} and $\rm\,I\hspace{-.1em}I\,$  in \Figref{fig:summary} and \Tabref{tab:2}. In Sect. \ref{sec:result}, we classified the results into four categories according to the classification of the flow and the accretion regimes (\Figref{fig:summary}).}

\subsection{Three-dimensional hydrodynamical simulations}\label{sec:hydro}
In this study, we performed nonisothermal 3D hydrodynamical simulations of the gas of the protoplanetary disk around a planet. Our simulations were performed in a spherical polar coordinate co-rotating with a planet with Athena++ \cite[]{White:2016,Stone:2020}. \AK{The computational domain ranges from $0$ to $\pi$ and $0$ to $2\pi$ in the polar and azimuthal directions, respectively.} Most of our methods of hydrodynamical simulations are the same as described in detail in \cite{Kurokawa:2018}, except for the configuration of the size of the inner boundary.  \AK{Since the initial condition is symmetrical in the vertical direction ($z$-direction), the structure of the planet-induced gas flow is symmetric with respect to the midplane. Our local simulations can not handle the gap opening. We focus on low-mass planets ($m\leq0.3$) which do not shape the global pressure gradient in both \citetalias{Kuwahara:2020} and $\rm\,I\hspace{-.1em}I\,$ (\Figref{fig:summary}). We discuss the case of high-mass planets in Sect. \ref{sec:high-mass}.}

\cite{Kurokawa:2018} fixed the size of the inner boundary for all of their simulations, but we varied it according to the mass of the planet. Following \citetalias{Kuwahara:2020}, assuming the density of the embedded planet $\rho_{\rm pl}=5$ g/cm$^{3}$ leads to the physical radius of the core, $R_{\rm pl}$, as given by
\begin{align}
R_{\rm pl}\simeq3\times10^{-3}m^{1/3}\,H\ \Biggl(\frac{\rho_{\rm pl}}{5\ {\rm g/}{\rm cm}^{3}}\Biggr)^{-1/3}\ \Biggl(\frac{M_{\ast}}{1\ M_{\odot}}\Biggr)^{1/3}\ \Biggl(\frac{a}{1\ {\rm au}}\Biggr)^{-1},\label{eq:planetaryradius}
\end{align}
where $M_{\ast}$, $M_{\odot}$ and $a$ are the stellar mass, the solar mass, and the orbital radius. We regard the size of $r_{\rm inn}$ as being determined by \Equref{eq:planetaryradius} with $a=1$ au.\footnote{In \citetalias{Kuwahara:2020}, we assumed $a=0.1$ au in \Equref{eq:planetaryradius}. However, the size of the Bondi radius is larger than the size of the physical radius of the planet when $m \lesssim0.005$ at $a=0.1$ au. This means that the planet does not have an envelope. To ensure that even low-mass planets ($m \lesssim0.005$) has an atmosphere, \AK{we assume $a=1$ au in this study. It ensures that $R_{\rm Bondi}>R_{\rm pl}$ for all of the parameter sets considered in our hydrodynamical simulations.}}

A planet is embedded in an inviscid gas disk and is orbiting around the central star at the distance $a$ with the orbital frequency $\Omega=\sqrt{GM_{\ast}/a^{3}}$. The unperturbed gas velocity in the local frame
\begin{align}
{\bm v}_{{\rm g}, \infty}(x)=\left(-\frac{3}{2}x-\mathcal{M}_{\rm hw}\right)\bm{e}_{y}.\label{eq:Keplerianshear}
\end{align}
From \Equref{eq:Keplerianshear}, the $x$-coordinate of the corotation radius for the gas can be described by
\begin{align}
    x_{g,\,{\rm cor}}=-\frac{2}{3}\mathcal{M}_{\rm hw}.\label{eq:corotation}
\end{align}
The Mach number of the headwind of the gas is defined as 
\begin{align}
\mathcal{M}_{\rm hw}=\frac{v_{\rm hw}}{c_{\rm s}},\label{eq:Mach}
\end{align}
where $v_{\rm hw}$ is the headwind of the gas,
\begin{align}
v_{\rm hw}=\eta v_{\rm K},\label{eq:vhw}
\end{align}
where 
\begin{align}
\eta =-\frac{1}{2}\left(\frac{c_{\rm s}}{v_{\rm K}}\right)^{2}\frac{\mathrm{d}\ln P}{\mathrm{d}\ln a}\label{eq:eta}
\end{align}
is a dimensionless quantity characterizing the pressure gradient of the disk gas, where $P$ is the pressure of the gas and $v_{\rm K}=a\Omega$ is the Kepler velocity. The disk gas rotates slower than Keplerian velocity due to the existence of the global pressure gradient. The Mach number of the headwind is $\mathcal{M}_{\rm hw}\simeq0.05\,(a/1\,{\rm au})^{1/4}$ in the MMSN model. In \citetalias{Kuwahara:2020}, we assumed $\mathcal{M}_{\rm hw}=0$ for all of hydrodynamical simulations. In this study, we assumed $\mathcal{M}_{\rm hw}=0.01,\,0.03$, and $0.1$.

We listed our parameter sets in \Tabref{tab:1}. The range of the dimensionless planetary masses, $m=0.003$--$0.03$, corresponds to planets of between three Moon masses and three Mars masses, $M_{\rm pl}=0.036$--0.36 $M_{\oplus}$, orbiting a solar-mass star at 1 au. 

\subsection{Three-dimensional orbital calculation of pebbles}\label{sec:orbital}
We calculated the trajectories of pebbles influenced by the planet-induced gas flow in the frame co-rotating with the planet (\Figref{fig:frame}). Most of our methods of orbital calculations of pebbles are the same as \citetalias{Kuwahara:2020}, except for the analysis of the headwind of the gas. 

In our co-rotating frame, the $x$- and $y$-component of the initial velocity of pebbles is given by the drift equations \cite[]{Weidenschilling:1977,Nakagawa:1986} 
\begin{align}
v_{x}&=-\frac{2\mathcal{M}_{\rm hw}{\rm St}}{1+{\rm St}^{2}},\label{eq:vx-int}\\
v_{y}&=-\frac{\mathcal{M}_{\rm hw}}{1+{\rm St}^{2}}-\frac{3}{2}x.\label{eq:vy-int}
\end{align} 
From \Equref{eq:vy-int}, the $x$-coordinate of the corotation radius for the pebble can be described by
\begin{align}
    x_{\rm peb,\,cor}=-\frac{2\mathcal{M}_{\rm hw}}{3\left(1+{\rm St}^2\right)},\label{eq:p-corotation}
\end{align}
where St is the dimensionless stopping time of a pebble, called the Stokes number, 
\AK{\begin{align}
{\rm St}=t_{\rm stop}\Omega.\label{eq:St}
\end{align}
}
We assumed ${\rm St}=10^{-4}$--$10^{0}$. The stopping time of the particle is described by
\begin{empheq}[left={t_{\rm stop}=\empheqlbrace}]{alignat=2}
&\displaystyle\frac{\rho_{\bullet}s}{\rho_{\rm g}c_{\rm s}}, \quad &(\text{Epstein regime}: s\leq\frac{9}{4}\lambda)\label{eq:Epstein-regime}\\
&\displaystyle\frac{4\rho_{\bullet}s^{2}}{9\rho_{\rm g}c_{\rm s}\lambda},  \quad &(\text{Stokes regime}: s\geq\frac{9}{4}\lambda)\label{eq:Stokes-regime}
\end{empheq}
where $\rho_{\bullet}$ is the internal density of the pebble, $s$ is the radius of the pebble, and $\lambda$ is the mean free path of the gas, $\lambda=\tilde{\mu} m_{\rm H}/\rho_{\rm g}\sigma_{\rm mol}=1.44\ {\rm cm}\ (a/{\rm 1\ au})^{11/4}$ with $\tilde{\mu}$, $m_{\rm H}$, and $\sigma_{\rm mol}$ being the mean molecular weight, $\tilde{\mu}=2.34$, the mass of the proton, and the molecular collision cross section, $\sigma_{\rm mol}=2\times10^{-15}\,{\rm cm}^{2}$ \cite[]{Chapman:1970,Weidenschilling:1977a,Nakagawa:1986}. The gas density at the midplane is given by $\rho_{\rm g}=\Sigma_{\rm g}/\sqrt{2\pi}H$, where $\Sigma_{\rm g}$ is the gas surface density, $\Sigma_{\rm g}=1700\,{\rm g\,cm}^{-2}\,(a/{\rm 1\,au})^{-3/2}$. \AK{The radius of a pebble is fixed in a orbital simulation, and the Stokes number is defined with the unperturbed gas density.}

We performed orbital calculations for three different settings:
\begin{enumerate}
\item 
Unperturbed flow case; hereafter \texttt{UP-mXX-hwYY} case\footnote{In \citetalias{Kuwahara:2020}, the unperturbed flow case is referred to as "Shear case".}, where we adopted unperturbed sub-Keplerian shear flow where the gas density is uniform. The \texttt{XX} and \texttt{YY} denote the adopted values of the planetary mass and the Mach number of the headwind. We omit \texttt{-mXX} or \texttt{-hwYY} when we do not specify the planetary mass or the Mach number of the headwind. Since the gas density around the planet is constant, there is no difference between the Stokes and the Epstein regimes in the case of unperturbed flow.
\item
Planet-induced flow case in the Epstein regime (\Equref{eq:Epstein-regime}); hereafter \texttt{PI-Epstein-mXX-hwYY} case. 
\item
Planet-induced flow case in the Stokes regime (\Equref{eq:Stokes-regime}); hereafter \texttt{PI-Stokes-mXX-hwYY} case. 
\end{enumerate}
In all of our hydrodynamical simulations, a hydrostatic envelope is formed around the planet. The density structure of an envelope is determined by the hydrostatic equilibrium with the gravity of the planet. The gas density increases significantly in the vicinity of the planet in the \texttt{PI} cases. Since the Stokes number in the Epstein regime is proportional to the reciprocal of the gas density, the effective Stokes number decreases as the gas density increases.\footnote{For simplicity, we do not consider the switch from the Epstein to the Stokes regime in the vicinity of the planet in both \citetalias{Kuwahara:2020} and this study. The mean free path of the gas becomes smaller as the gas density increases. It may lead to a switch of the drag law in the region very close to the planet.} For the latter two cases, we switched the gas flow from unperturbed sub-Keplerian to planet-induced gas flow obtained by hydrodynamical simulations at $r=r_{\rm out}$. We used the final state of the hydro-simulations data $(t=t_{\rm end})$, where the flow field seems to have reached a steady state. We interpolated the gas velocity using the bilinear interpolation method \citepalias[see Appendix B in][]{Kuwahara:2020}.

When $m=0.01$ and $0.03$, we found that the horseshoe flow formed unexpected vortices, which influence the pebble trajectories. The origin of these vortices is unknown, but it is likely to be a numerical artifact due to the spherical polar coordinates centered at the planet, in which the resolution becomes too low to resolve the horseshoe flow far from the planet when the assumed planet mass is small. In the same manner as in \citetalias{Kuwahara:2020}, only in this case do we use the limited part of the calculation domain, $r\leq0.6r_{\rm out}$, to avoid the effects of the vortices.

\subsection{Calculation of accretion probability of pebbles}
\subsubsection{Width of accretion window and accretion cross section}
We defined the width of the accretion window as 
\begin{align}
w_{\rm acc}(z)=\sum_{i}\left(x_{{\rm max}, i}(z)-x_{{\rm min}, i}(z)\right),\label{eq:wacc}
\end{align}
where $x_{{\rm max}, i}(z)$ and $x_{{\rm min}, i}(z)$ are the maximum and minimum values of the $x$-component of the starting point of accreted pebbles at a certain height. The subscript, $i$, denotes the number of the accretion window. When we include the headwind of the gas, accretion of pebbles occurs asymmetrically with respect to the corotation radius for the pebble. In the unperturbed flow, the width of the accretion window is identical to the maximum impact parameter of accreted pebbles, $b_{x}$, when ${\rm St}<1$ as $x_{{\rm min}, i}(z)=0$ (Eqs. \ref{eq:b-hw} and \ref{eq:b-sh}). Using this definition, we defined the accretion cross section as 
\begin{align}
A_{\rm acc}=\sum_{i}\left(\int_{z_{{\rm min}, i}}^{z_{{\rm max}, i}}\int_{x_{{\rm min}, i}(z)}^{x_{{\rm max}, i}(z)}\mathrm{d}x\mathrm{d}z\right),\label{eq:Aacc}
\end{align}
where $z_{{\rm max}, i}$ and $z_{{\rm min}, i}$ are the maximum and the minimum value of the $z$-component of the starting point of accreted pebbles. We reduced the spatial intervals stepwise near the edge of the accretion window, and determined $x_{{\rm max}, i}(z)$ and $x_{{\rm min}, i}(z)$ with sufficient
accuracy. The initial spatial intervals are $10^{-2}\times \min\left(b_{x,{\rm hw}}, b_{x,{\rm sh}}\right)$ in the $x$ and $z$ directions, where $b_{x,\,{\rm hw}}$ and $b_{x,\,{\rm sh}}$ are the maximum impact parameter of accreted pebbles in the unperturbed flow (Eqs. \ref{eq:b-hw} and \ref{eq:b-sh}). 

\subsubsection{Accretion probability}
We define the accretion probability of pebbles as
\begin{align}
P_{\rm acc}=\frac{\dot{M}_{\rm p}}{\dot{M}_{\rm disk}},\label{eq:Pacc}
\end{align}
where $\dot{M}_{\rm p}$ is the accretion rate of pebbles onto a protoplanet and $\dot{M}_{\rm disk}$ is the radial inward mass flux of pebbles in the gas disk described by 
\begin{align}
\dot{M}_{\rm disk}=2\pi a\Sigma_{\rm p}|v_{r}|,\label{eq:M-dot-disk}
\end{align}
where $\Sigma_{\rm p}$ is the surface density of pebbles.
The density distribution of pebbles is described by
\begin{align}
\rho_{\rm p}(z)=\frac{\Sigma_{\rm p}}{\sqrt{2\pi}H_{\rm p}}\exp\left[-\frac{1}{2}\left(\frac{z}{H_{\rm p}}\right)^{2}\right],\label{eq:pebble-density}
\end{align}
where $H_{\rm p}$ is the scale height of pebbles \cite[]{Durbrulle:1995,Cuzzi:1993,Youdin:2007}:
\begin{align}
H_{\rm p}=\left(1+\frac{{\rm St}}{\alpha}\frac{1+2{\rm St}}{1+{\rm St}}\right)^{-1/2}H,\label{eq:pebble-scaleheight}
\end{align}
where $\alpha$ is the dimensionless turbulent parameter in the disk introduced by \cite{Shakura:1973}. Our calculation of accretion probability assumed that pebbles have a vertical distribution given by \Equref{eq:pebble-density}. This approach neglects the effect of random motion of individual particles \citepalias[see,][for the discussion]{Kuwahara:2020}. The accretion rate of pebbles, $\dot{M}_{\rm p}$, is divided into two formulas:
\begin{align}
\dot{M}_{\rm p, 2D}=\sum_{i}\left(\int_{x_{{\rm min}, i}(0)}^{x_{{\rm max}, i}(0)}\Sigma_{\rm p}\bm{v}_{{\rm p},\infty}\mathrm{d}x\right),\label{eq:M-dot-2D-sim}
\end{align}
in the 2D case, and
\begin{align}
\dot{M}_{\rm p, 3D}=\sum_{i}\left(2\int_{z_{{\rm min}, i}}^{z_{{\rm max}, i}}\int_{x_{{\rm min}, i}(z)}^{x_{{\rm max}, i}(z)}\rho_{{\rm p},\infty}(z)\bm{v}_{{\rm p},\infty}\mathrm{d}x\mathrm{d}z\right),\label{eq:M-dot-sim}
\end{align}
in the 3D case. In order to account for the accretion from $z<0$, we multiply \Equref{eq:M-dot-sim} by two. The accretion probabilities for a fixed dimensionless planetary mass, $m$, in both 2D and 3D do not depend on the orbital radius, $a$ \citepalias[see Appendix C in][]{Kuwahara:2020}. Following \citetalias{Kuwahara:2020}, we fixed the inward pebble mass flux as $\dot{M}_{\rm disk}=10^{2}\,M_{\oplus}$/Myr, which is consistent with the typical value of the pebble flux used in a previous study \cite[]{Lambrechts:2019}. 

\section{Results}\label{sec:result}
\iffigure
 \begin{figure}[htbp]
 \resizebox{\hsize}{!}
 {\includegraphics{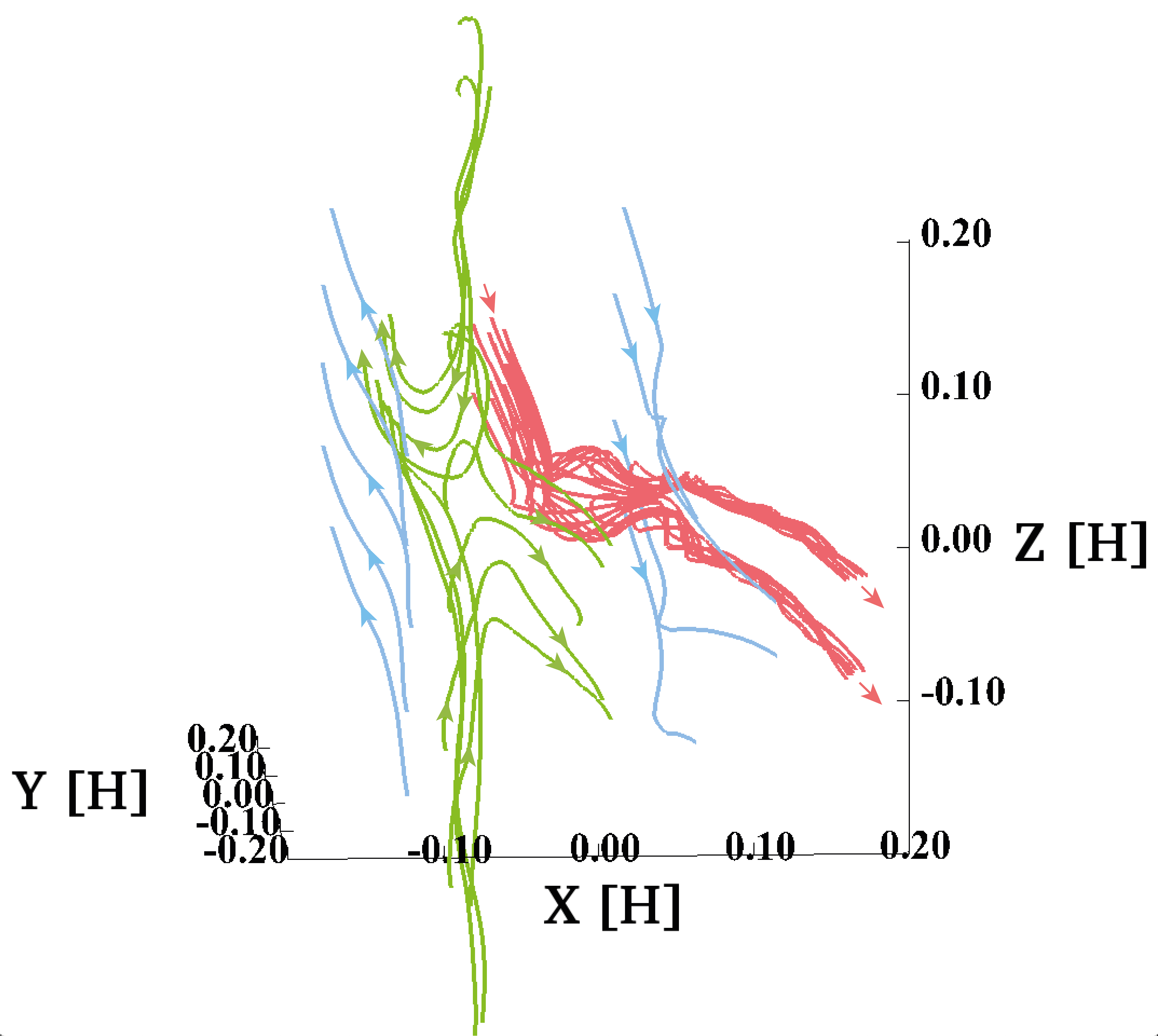}} 
 \caption{Streamlines of 3D planet-induced gas flow around the planet. The result obtained from \texttt{m003-hw01} at $t=50$. The red, green, and blue solid lines are the recycling streamlines, the horseshoe streamlines, and the Keplerian shear streamlines, respectively. \AK{For the recycling streamlines, we only plot the streamlines which pass over the surface of the Bondi sphere. The arrows represent the direction of the gas flow.}}
\label{fig:gasflow-3D}
\end{figure}

\fi
\iffigure
 \begin{figure}[htbp]
 \resizebox{\hsize}{!}
 {\includegraphics{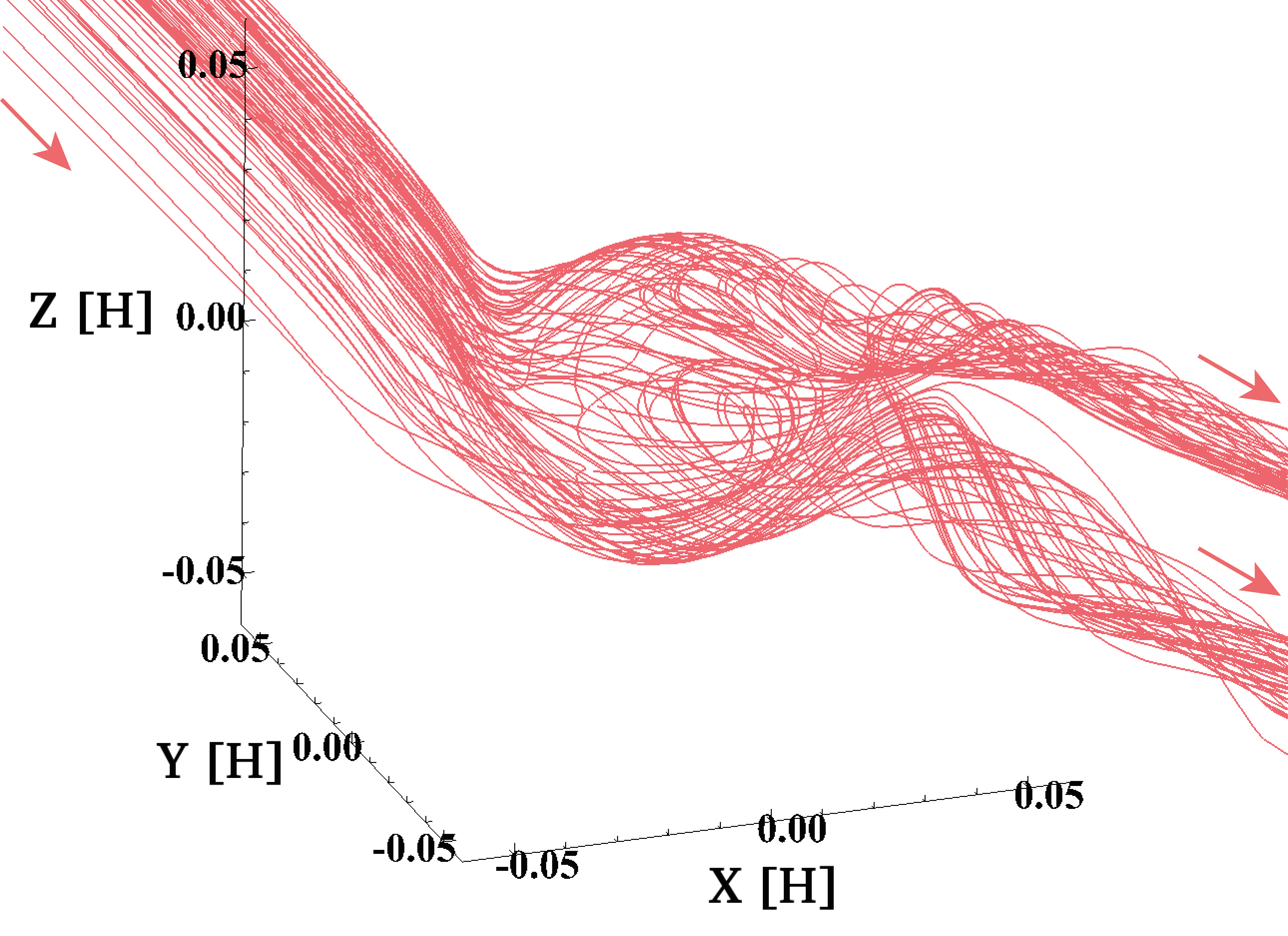}} 
 \caption{Enlarged view of \Figref{fig:gasflow-3D}. We only plot the recycling streamlines.}
\label{fig:gasflow-3D-2}
\end{figure}
\fi

\iffigure
 \begin{figure*}[htbp]
 \resizebox{\hsize}{!}
 {\includegraphics{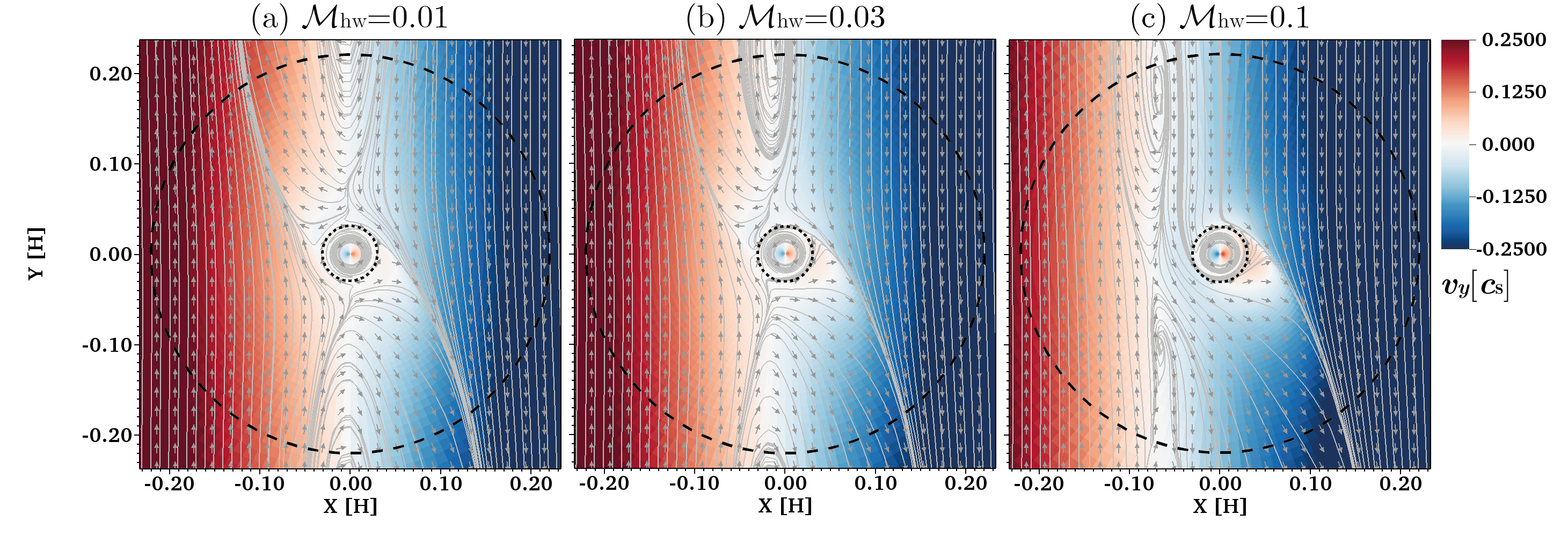}} 
 \caption{Flow structure around a planet with $m=0.03$ at the midplane of the disk. \textit{Panel a}: result obtained from \texttt{m003-hw001} at $t=50$. \textit{Panel b}: result obtained from \texttt{m003-hw003} at $t=50$. \textit{Panel c}: result obtained from \texttt{m003-hw01} at $t=50$. Color contour represents the flow speed in the $y$-direction. The vertical and horizontal axis are normalized by the scale height of the disk. The solid lines correspond to the specific streamlines. The dotted and dashed lines are the Bondi and Hill radius of the planet, respectively. We note that the length of the arrows does not scale with the flow speed.}
\label{fig:gasflow-2D}
\end{figure*}
\fi
\iffigure
 \begin{figure}[htp]
 \resizebox{\hsize}{!}
 {\includegraphics{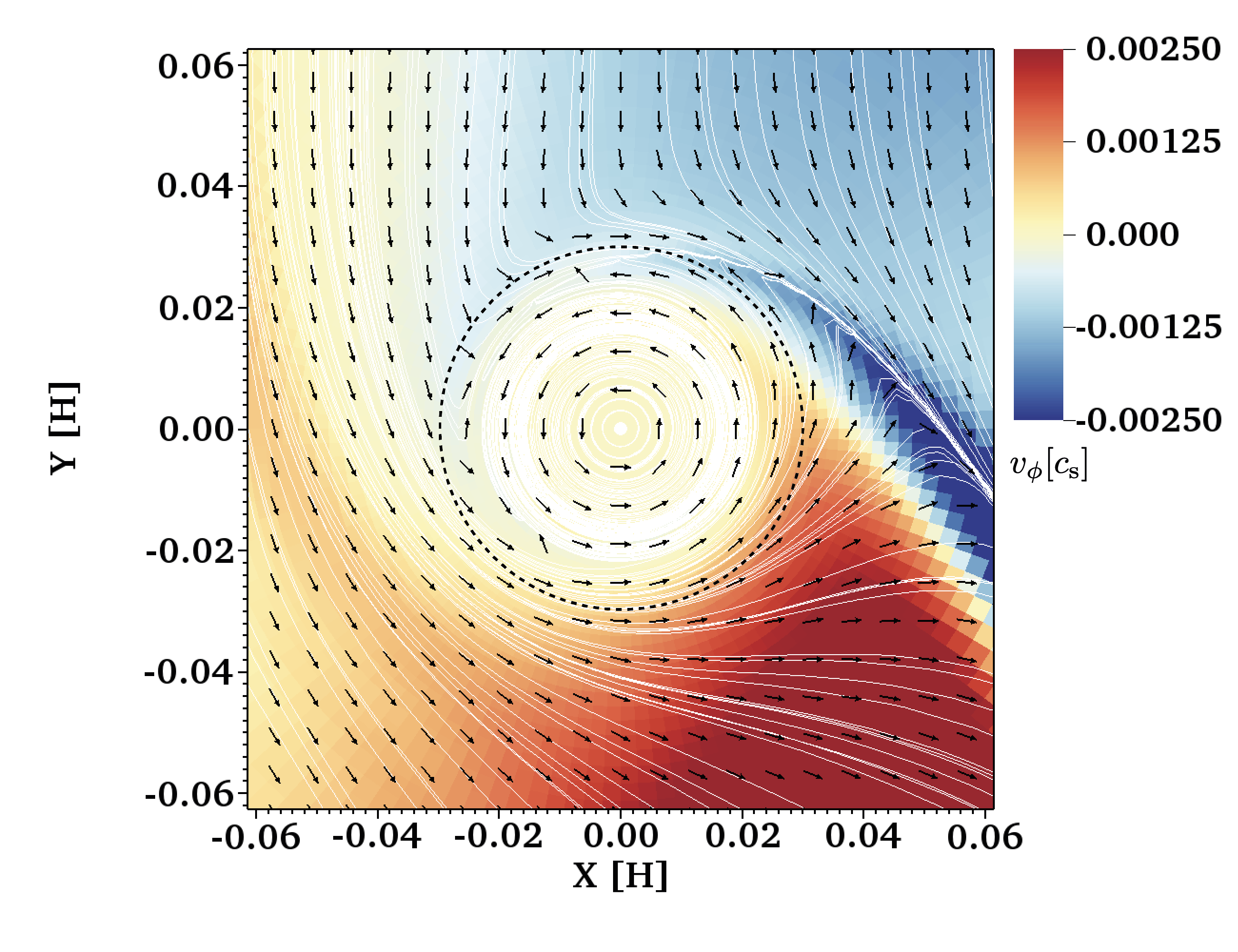}} 
 \caption{Enlarged view of \Figref{fig:gasflow-2D}c, but color contour represents the flow speed in the azimuthal direction.}
\label{fig:v-phi}
\end{figure}
\fi
\iffigure
 \begin{figure*}[htbp]
 \resizebox{\hsize}{!}
 {\includegraphics{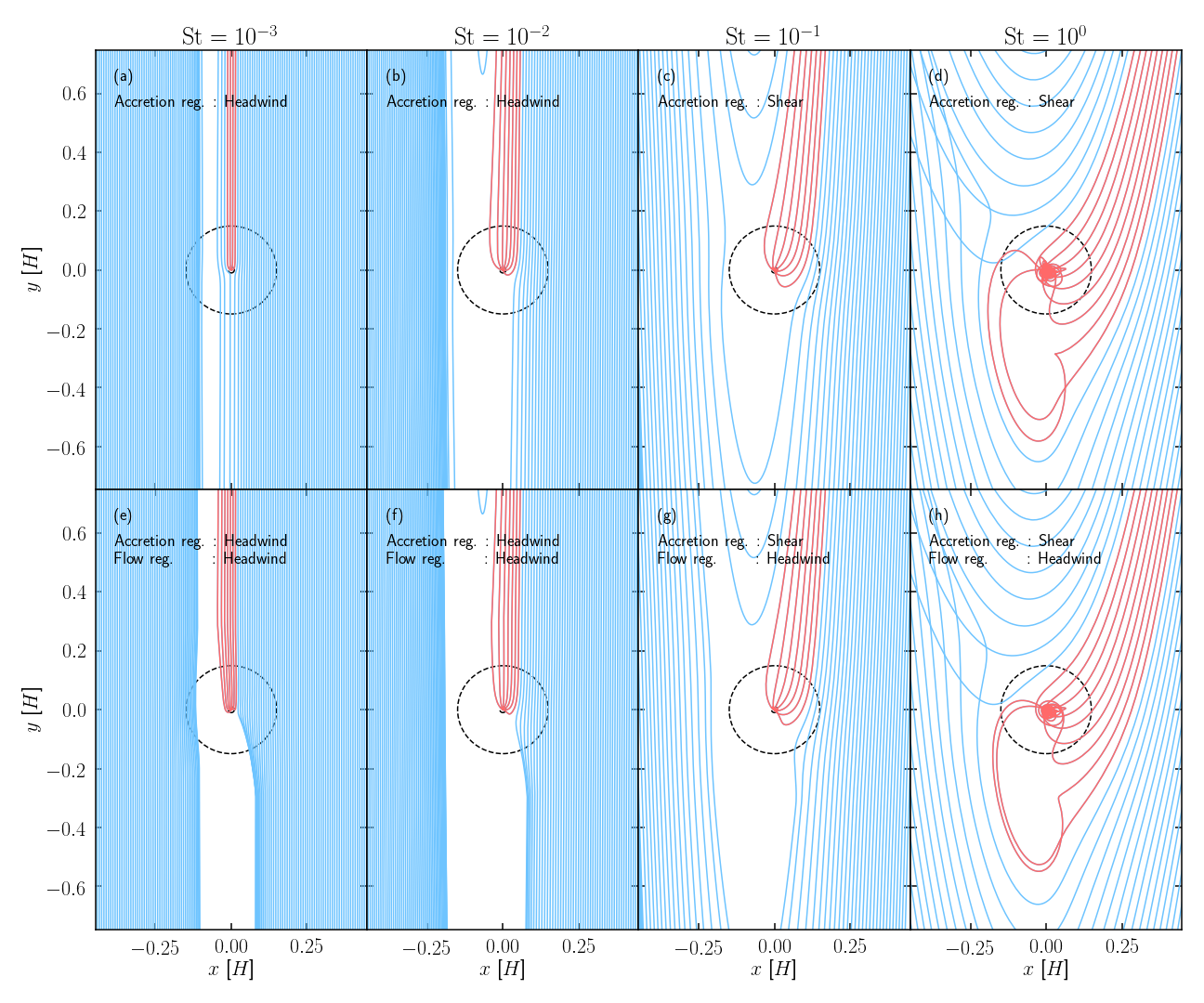}} 
 \caption{Trajectories of pebbles in \texttt{UP-m001-hw01} case (\textit{top}) and \texttt{PI-Stokes-m001-hw01} case (\textit{bottom}) with different Stokes numbers at the midplane of the disk. We set $z_{\rm s}=0$ for all cases. The red and blue solid lines correspond to the trajectories of pebbles which accreted and did not accrete onto the planet, respectively. The dashed circles show the Hill radius of the planet. The sizes of the Hill and Bondi radii are $0.15\,[H]$ and $0.01\,[H]$. The black dot at the center of each panel denotes the position of the planet. The interval of pebbles at their initial locations is $0.05$ [$H$]. The regimes of pebble accretion and the planet-induced gas flow are determined by Eqs. (\ref{eq:pebble-transition}) and (\ref{eq:m-flow-2}).}
\label{fig:peb-line1}
\end{figure*}
\fi

\iffigure
 \begin{figure}[htbp]
 \resizebox{\hsize}{!}
 {\includegraphics{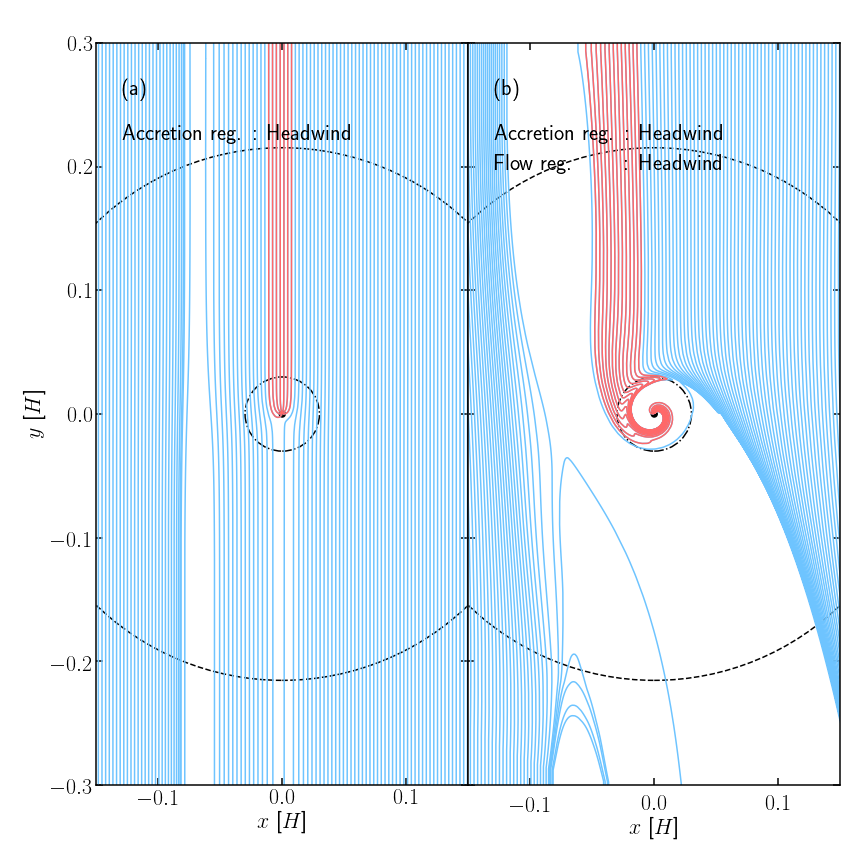}} 
 \caption{Trajectories of pebbles in \texttt{UP-m003-hw01} case (\textit{left panel}) and \texttt{PI-Stokes-m003-hw01} case (\textit{right panel}) with ${\rm St}=10^{-4}$ at the midplane of the disk. We set $z_{\rm s}=0$. The red and blue solid lines correspond to the trajectories of pebbles which accreted and did not accrete onto the planet, respectively. The dotted-dashed and dashed circles show the Bondi and the Hill radius of the planet, respectively. The sizes of the Bondi radius and the Hill radius are $0.03\,[H]$ and $0.22\,[H]$. The black dots at the center of each panel denote the position of the planet. The interval of pebbles at their initial locations is $0.003$ [$H$].}
\label{fig:peb-line2}
\end{figure}
\fi
\iffigure
 \begin{figure}[htbp]
 \resizebox{\hsize}{!}
 {\includegraphics{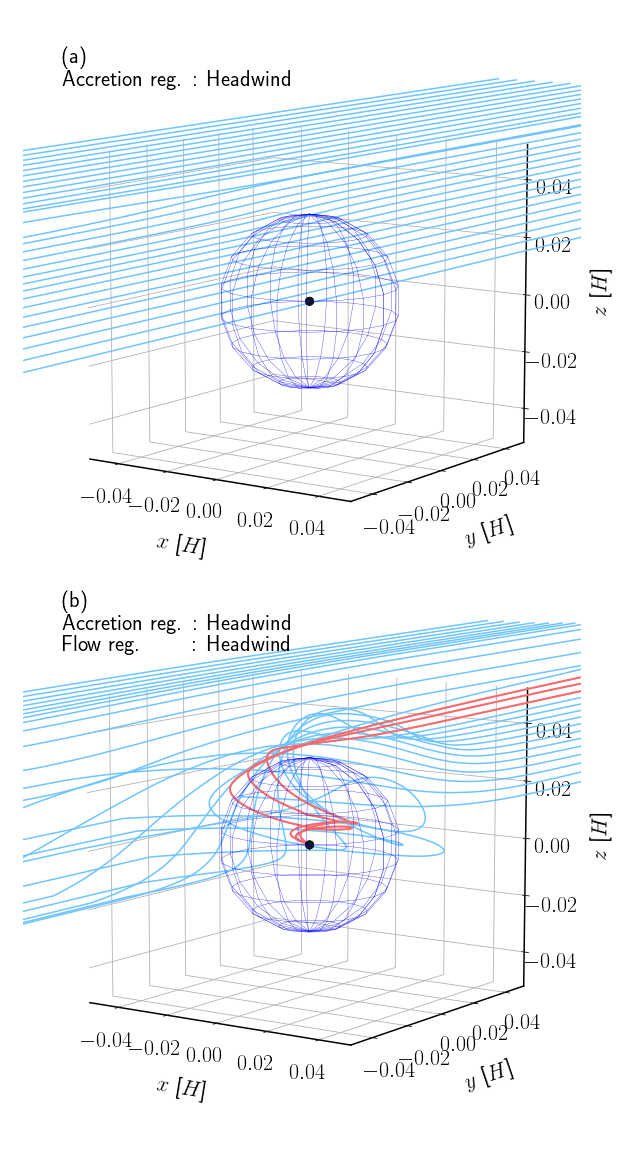}} 
 \caption{Trajectories of pebbles with ${\rm St}=10^{-4}$ in \texttt{UP-m003-hw01} case (\textit{top}) and \texttt{PI-Stokes-m003-hw01} case (\textit{bottom}). The height of the initial position of the pebbles is $z_{\rm s}=0.03$ [$H$]. The red and blue solid lines correspond to the trajectories of pebbles which accreted and did not accrete onto the planet, respectively. The black dots at the center of each panel denote the position of the planet. The sphere around the planet represents the Bondi radius of the planet. We only plot the trajectories within the region where $r<2R_{\rm Hill}$. The interval of pebbles at their initial locations is $0.01$ [$H$].}
\label{fig:peb-line3}
\end{figure}
\fi
\iffigure
 \begin{figure}[htbp]
 \resizebox{\hsize}{!}
 {\includegraphics{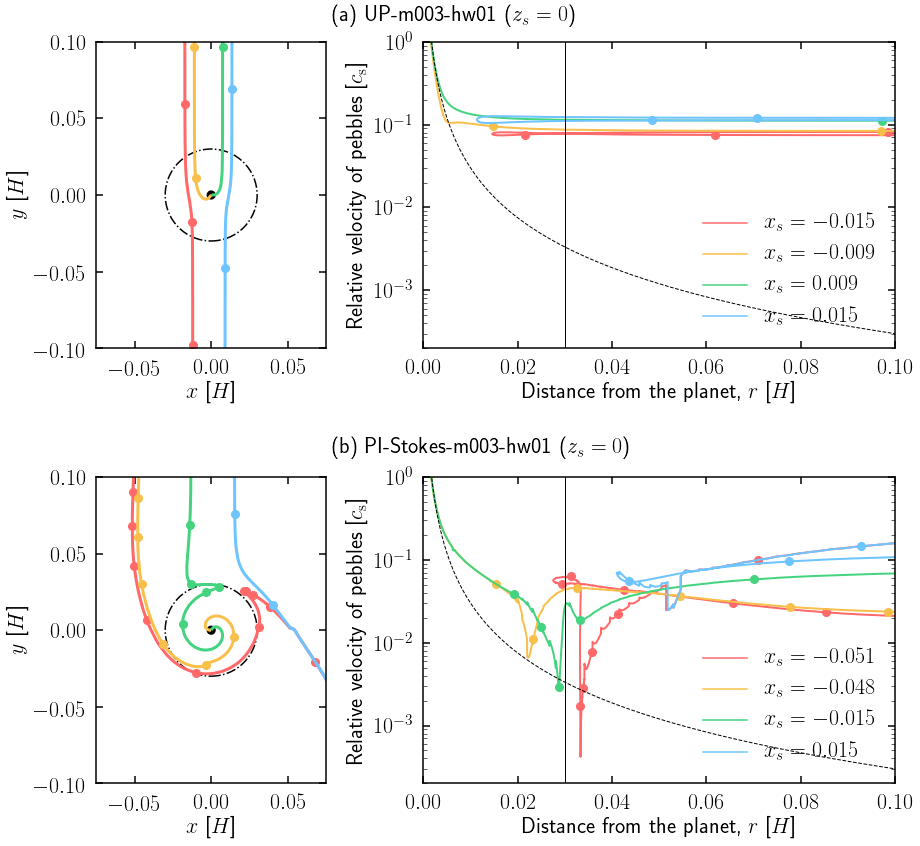}} 
 \caption{Trajectories (\textit{left}) and the relative velocity of pebbles to the planet (\textit{right}) with ${\rm St}=10^{-4}$. \textit{Panel a}: results obtained from \texttt{UP-m003-hw01} case. \textit{Panel b}: results obtained from \texttt{PI-Stokes-m003-hw01} case. We set $z_{\rm s}=0$ [$H$]. Different colors correspond to different $x_{\rm s}$ as indicated in figure legends. The dots on the solid lines mark intervals of $\Omega^{-1}$. The black dot and the dotted-dashed circle in the left column show the position of the planet and the  Bondi radius of the planet. The vertical solid and dashed lines in the right column show the size of the Bondi radius and the terminal velocity of pebbles (\Equref{eq:terminal}).}
\label{fig:v-peb1}
\end{figure}
\fi

\iffigure
 \begin{figure}[htp]
 \resizebox{\hsize}{!}
 {\includegraphics{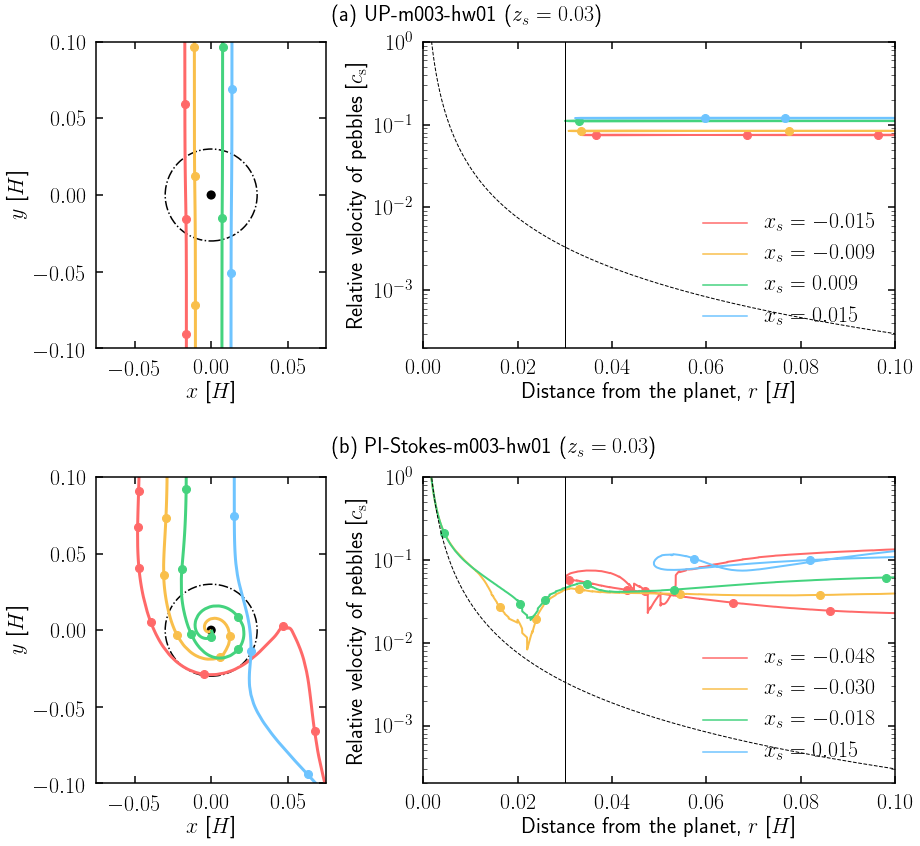}} 
 \caption{Same as \Figref{fig:v-peb1}, but we set $z_{\rm s}=0.03$ [$H$].}
\label{fig:v-peb2}
\end{figure}
\fi

\iffigure
 \begin{figure*}[htbp]
 \resizebox{\hsize}{!}
 {\includegraphics{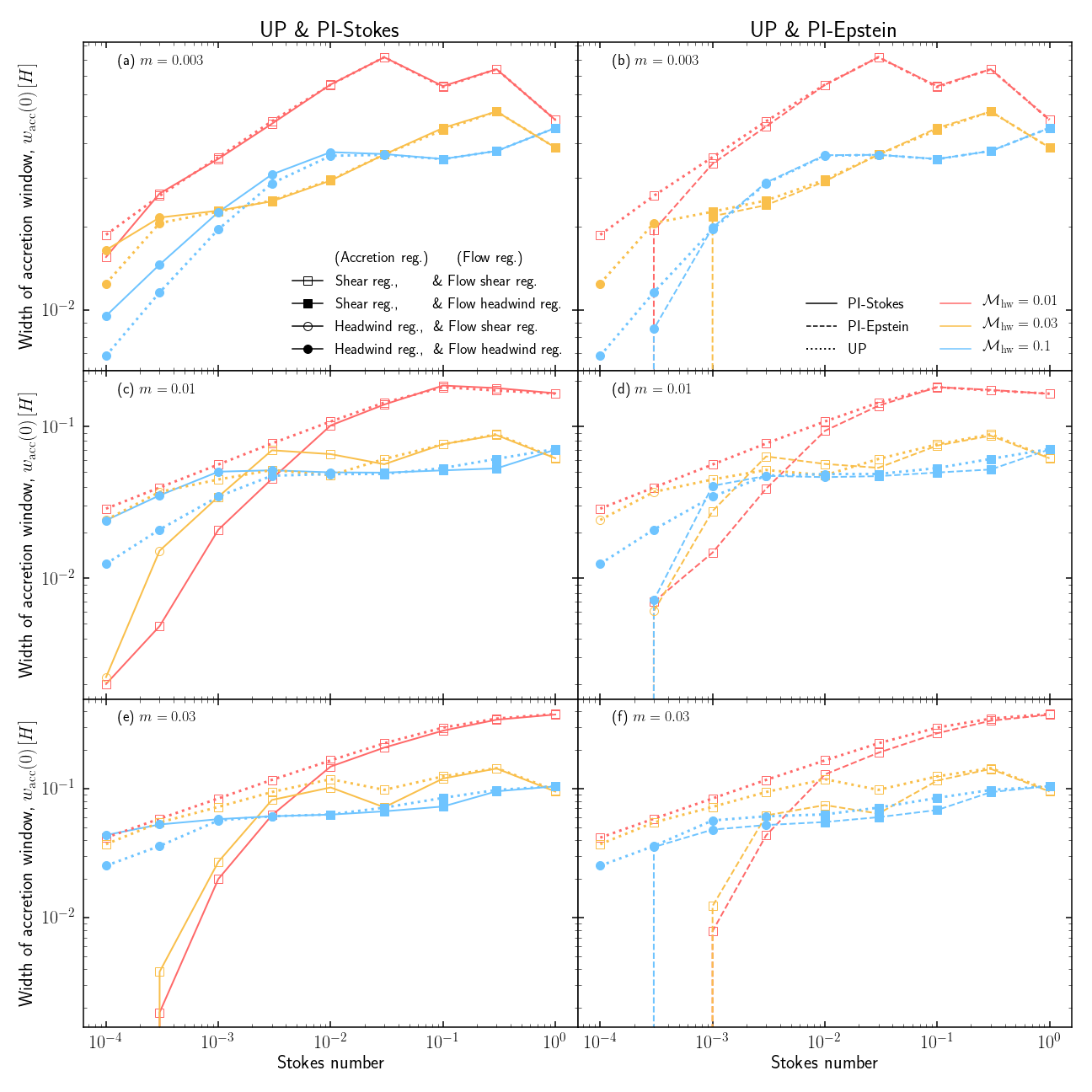}} 
 \caption{Width of accretion window in the midplane, $w_{\rm acc}(0)$, as a function of the Stokes number in the UP (dotted lines), \texttt{PI-Stokes} (solid lines), and the \texttt{PI-Epstein} cases (dashed lines). Left column compares the results between \texttt{UP} and \texttt{PI-Stokes} cases. Right column compares the results between \texttt{UP} and \texttt{PI-Epstein} cases. The masses of the planet from top to bottom rows are $m=0.003$, $0.01$, and $0.03$, respectively. Colors indicate the Mach number of the headwind of the gas: $\mathcal{M}_{\rm hw}=0.01$ (red), $\mathcal{M}_{\rm hw}=0.03$ (yellow), and  $\mathcal{M}_{\rm hw}=0.1$ (blue). The open and filled squares and circles denote the regimes of pebble accretion and the planet-induced gas flow at the given parameters.}
\label{fig:w-acc}
\end{figure*}
\fi
\iffigure
 \begin{figure*}[htbp]
 \resizebox{\hsize}{!}
 {\includegraphics{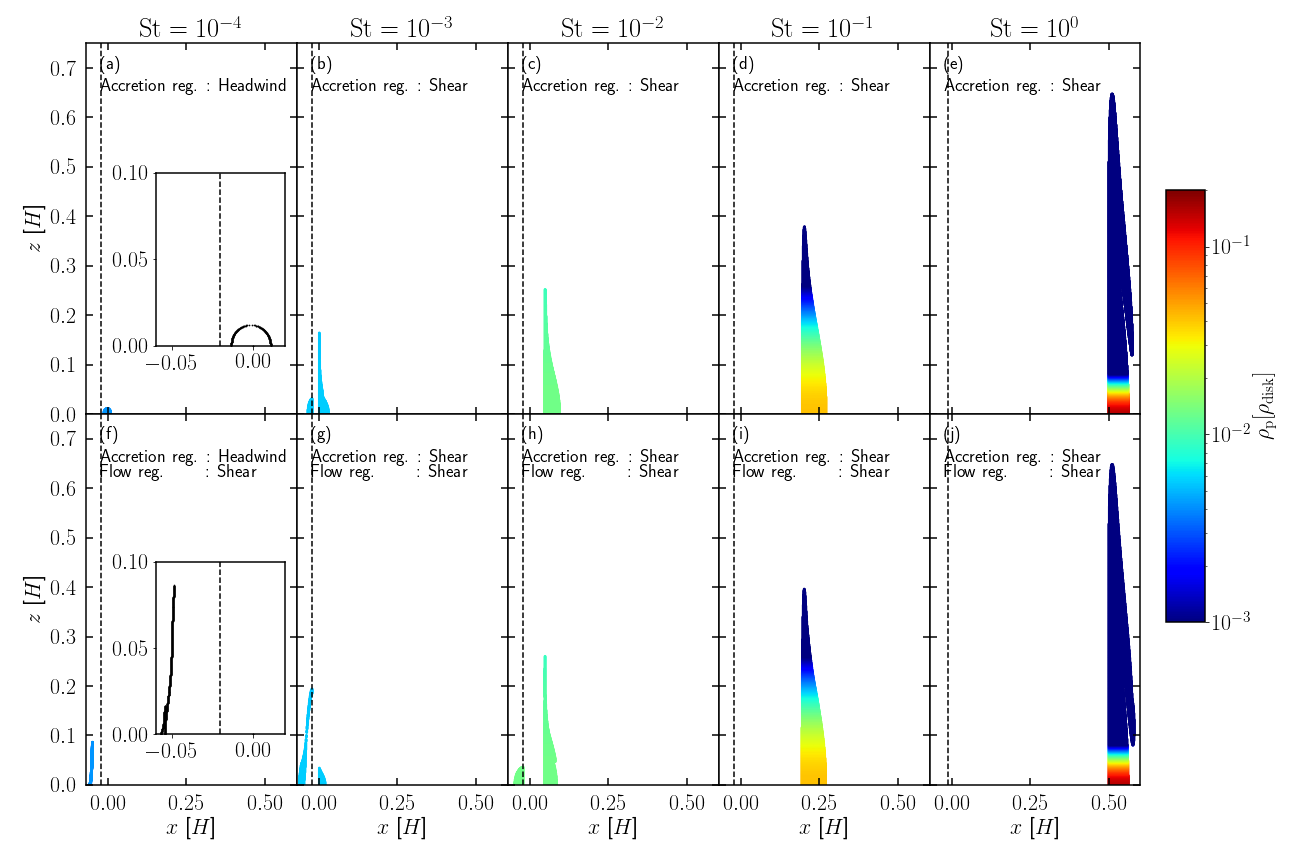}} 
 \caption{Accretion window with the different Stokes numbers in the \texttt{UP-m001-hw003} (\textit{top}) and the \texttt{PI-Stokes-m001-hw003} cases (\textit{bottom}). We assumed $\alpha=10^{-3}$ and the dust-to-gas ratio is equal to $10^{-2}$. Color represents the density of the pebbles expressed by \Equref{eq:pebble-density} normalized by the gas density. The vertical dashed lines correspond to the corotation radius for pebbles (\Equref{eq:p-corotation}). The two panels in \textit{panels a and e} show the enlarged outlines of accretion windows. We note that the color contour is saturated for the $\rho_{\rm p}\lesssim10^{-3}$.}
\label{fig:Aacc-m001-hw003}
\end{figure*}
\fi
\iffigure
 \begin{figure*}[htbp]
 \resizebox{\hsize}{!}
 {\includegraphics{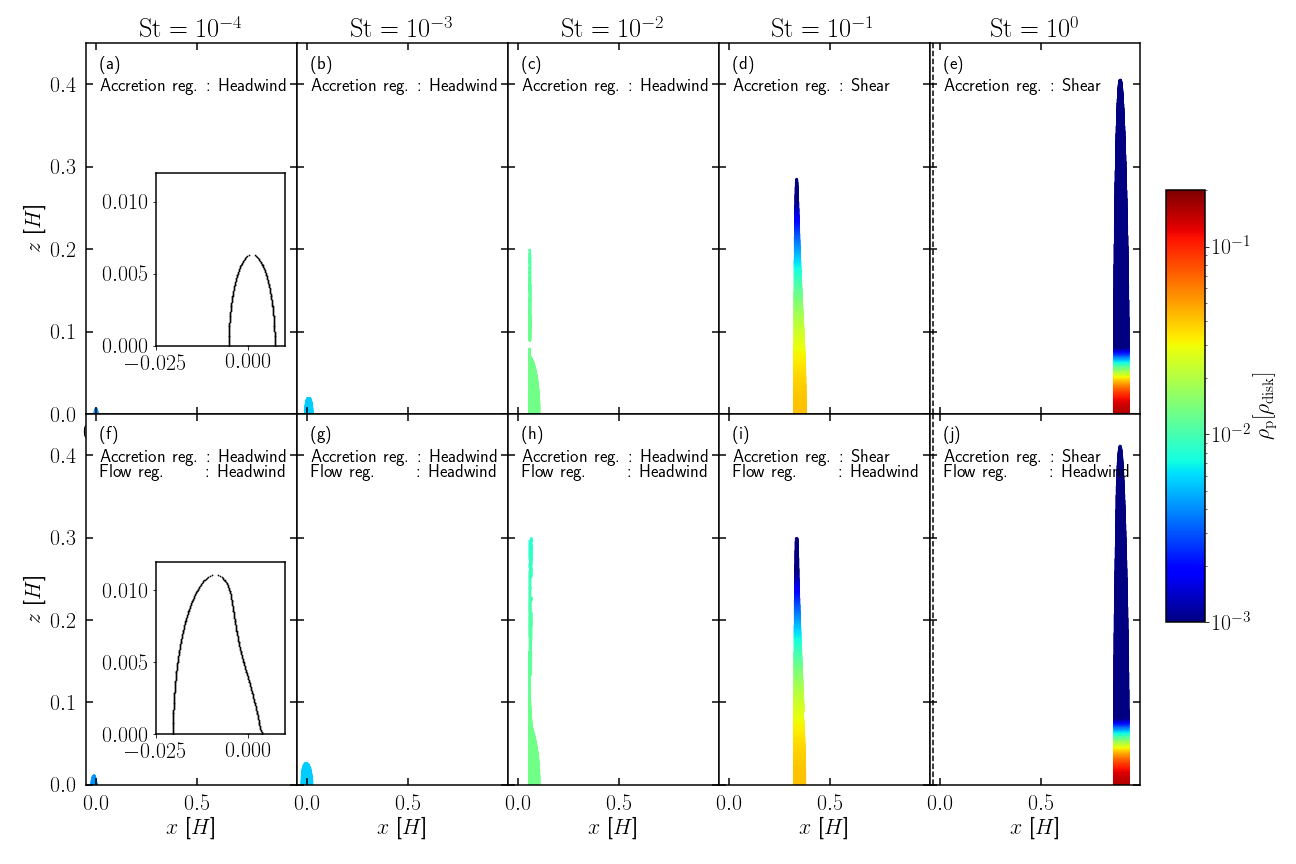}} 
 \caption{Same as \Figref{fig:Aacc-m001-hw003}, but in the \texttt{UP-m001-hw01} case (\textit{top}) and the \texttt{PI-Stokes-m001-hw01} case (\textit{bottom}).}
\label{fig:Aacc-m001-hw01}
\end{figure*}
\fi
\iffigure
 \begin{figure*}[htbp]
 \resizebox{\hsize}{!}
 {\includegraphics{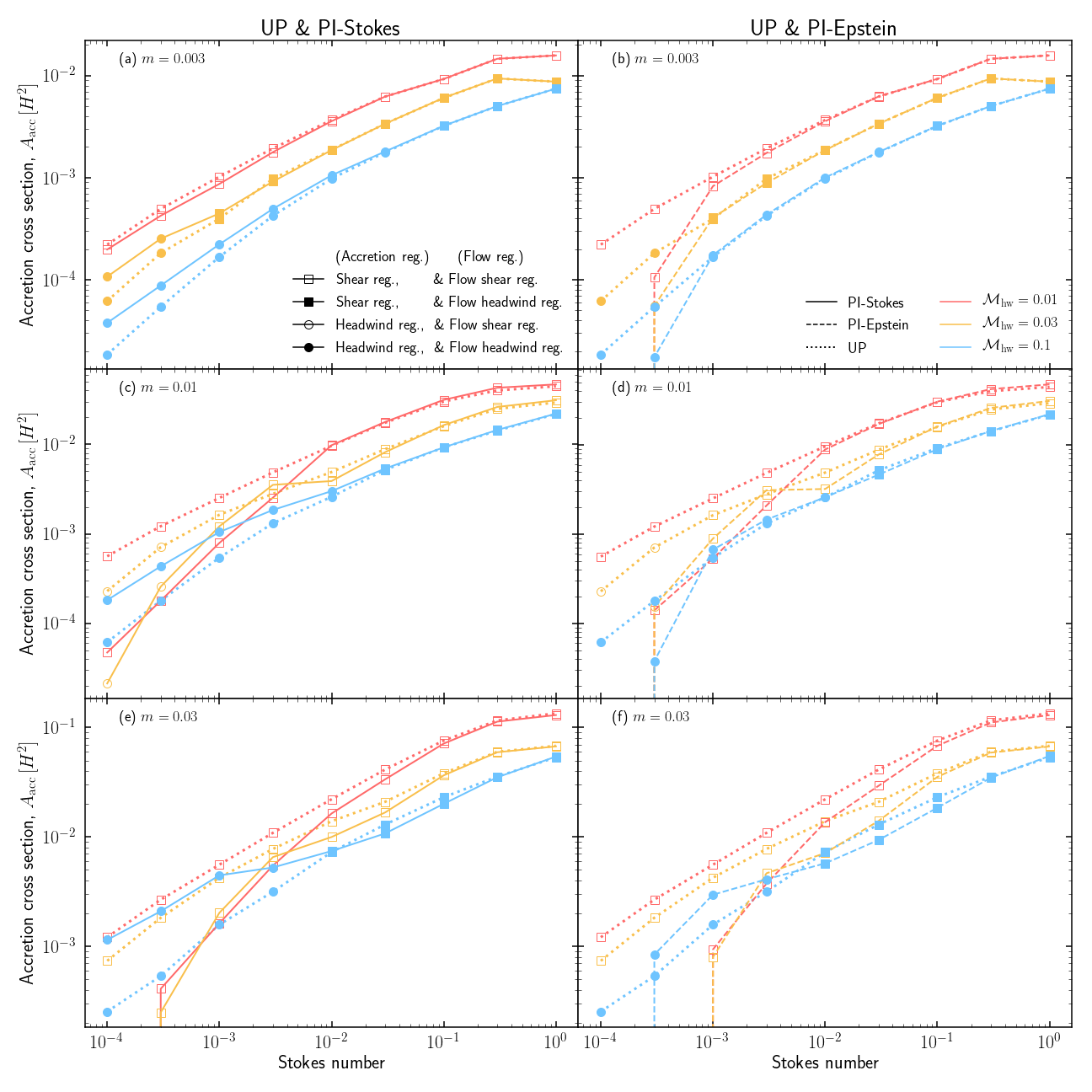}} 
 \caption{Accretion cross section, $A_{\rm acc}$, as a function of the Stokes number in \texttt{UP} (dotted lines), \texttt{PI-Stokes} (solid lines), and \texttt{PI-Epstein} cases (dashed lines). Left column compares the results  between \texttt{UP} and \texttt{PI-Stokes} cases. Right column compares the results between \texttt{UP} and \texttt{PI-Epstein} cases. The masses of the planet from top to bottom rows are $m=0.003$, $0.01$, and $0.03$, respectively. Colors indicate the Mach number of the headwind of the gas: $\mathcal{M}_{\rm hw}=0.01$ (red), $\mathcal{M}_{\rm hw}=0.03$ (yellow), and  $\mathcal{M}_{\rm hw}=0.1$ (blue). The open and filled squares and circles denote the regimes of pebble accretion and the planet-induced gas flow at the given parameters.}
\label{fig:A-acc}
\end{figure*}
\fi
\iffigure
 \begin{figure*}[htbp]
 \resizebox{\hsize}{!}
 {\includegraphics{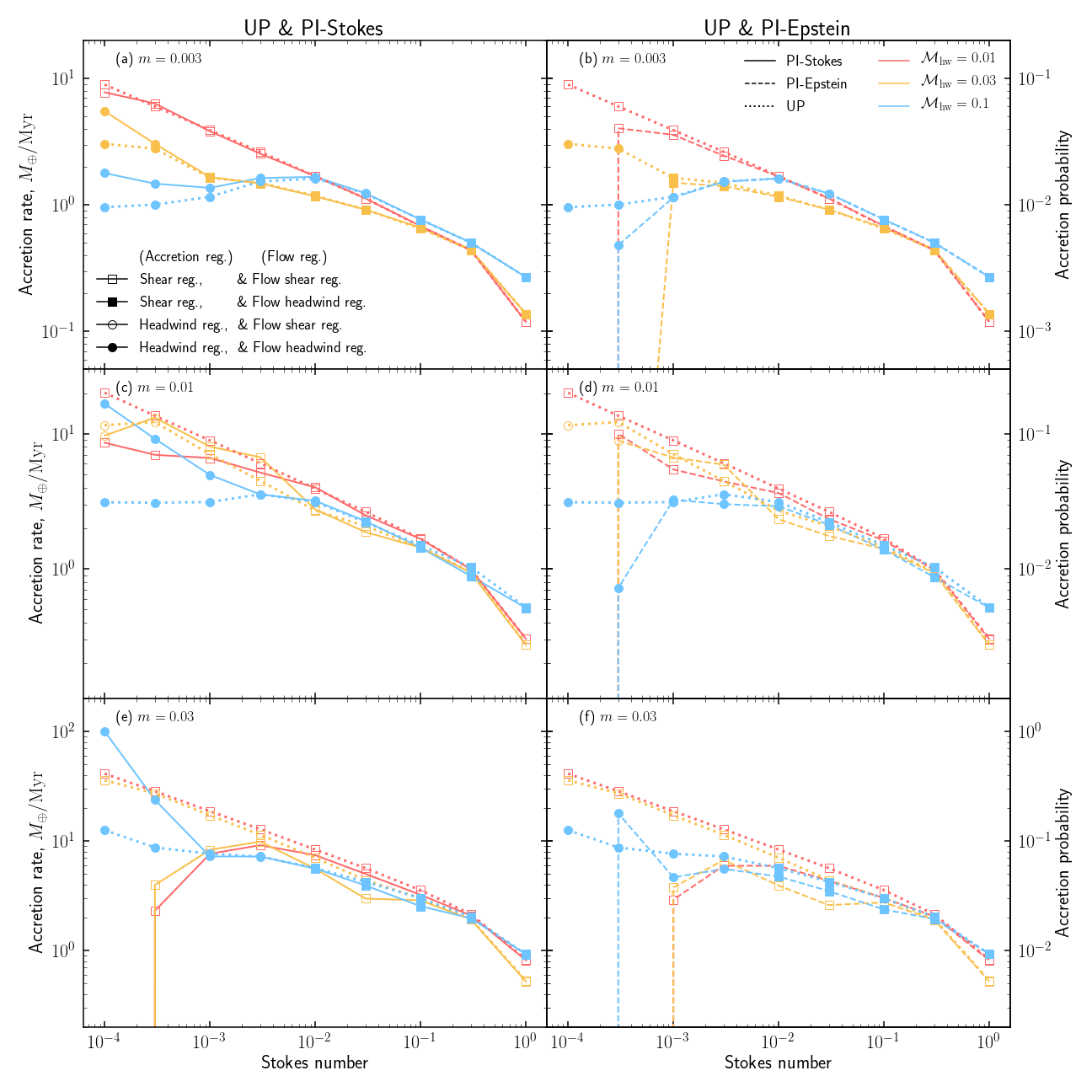}} 
 \caption{Two-dimensional accretion rate, $\dot{M}_{\rm 2D}$, (left vertical axis) and probability (right vertical axis) as a function of the Stokes number in \texttt{UP} (dotted lines), \texttt{PI-Stokes} (solid lines), and \texttt{PI-Epstein} cases (dashed lines). Left column compares the results between \texttt{UP} and \texttt{PI-Stokes} cases. Right column compares the results between \texttt{UP} and \texttt{PI-Epstein} cases. The masses of the planet from top to bottom rows are $m=0.003$, $0.01$, and $0.03$, respectively. Colors indicate the Mach number of the headwind of the gas: $\mathcal{M}_{\rm hw}=0.01$ (red), $\mathcal{M}_{\rm hw}=0.03$ (yellow), and  $\mathcal{M}_{\rm hw}=0.1$ (blue). The open and filled squares and circles denote the regimes of pebble accretion and the planet-induced gas flow at the given parameters.}
\label{fig:Mdot-2D}
\end{figure*}
\fi

\iffigure
 \begin{figure*}[htbp]
 \resizebox{\hsize}{!}
 {\includegraphics{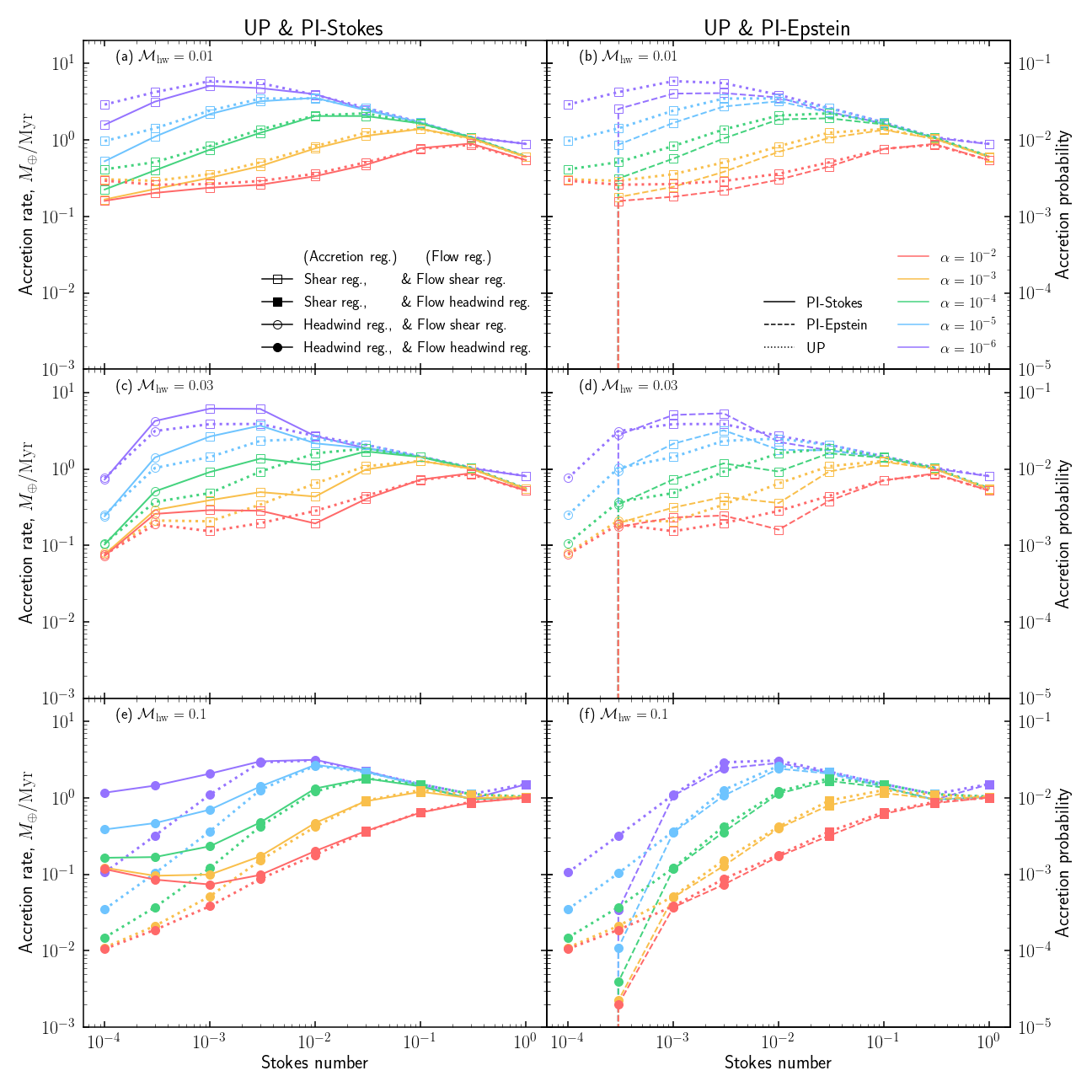}} 
 \caption{Three-dimensional accretion rate, $\dot{M}_{\rm 3D}$, (left vertical axis) and probability (right vertical axis) as a function of the Stokes number in \texttt{UP-m001} (dotted lines), \texttt{PI-Stokes-m001} (solid lines), and \texttt{PI-Epstein-m001} cases (dashed lines). Left column compares the results between \texttt{UP} and \texttt{PI-Stokes} cases. Right column compares the results between \texttt{UP} and \texttt{PI-Epstein} cases. Colors indicate the turbulent parameter, $\alpha$. The open and filled squares and circles denote the regimes of pebble accretion and the planet-induced gas flow at the given parameters.}
\label{fig:Mdot-3D-m001}
\end{figure*}
\fi
\iffigure
 \begin{figure*}[htbp]
 \resizebox{\hsize}{!}
 {\includegraphics{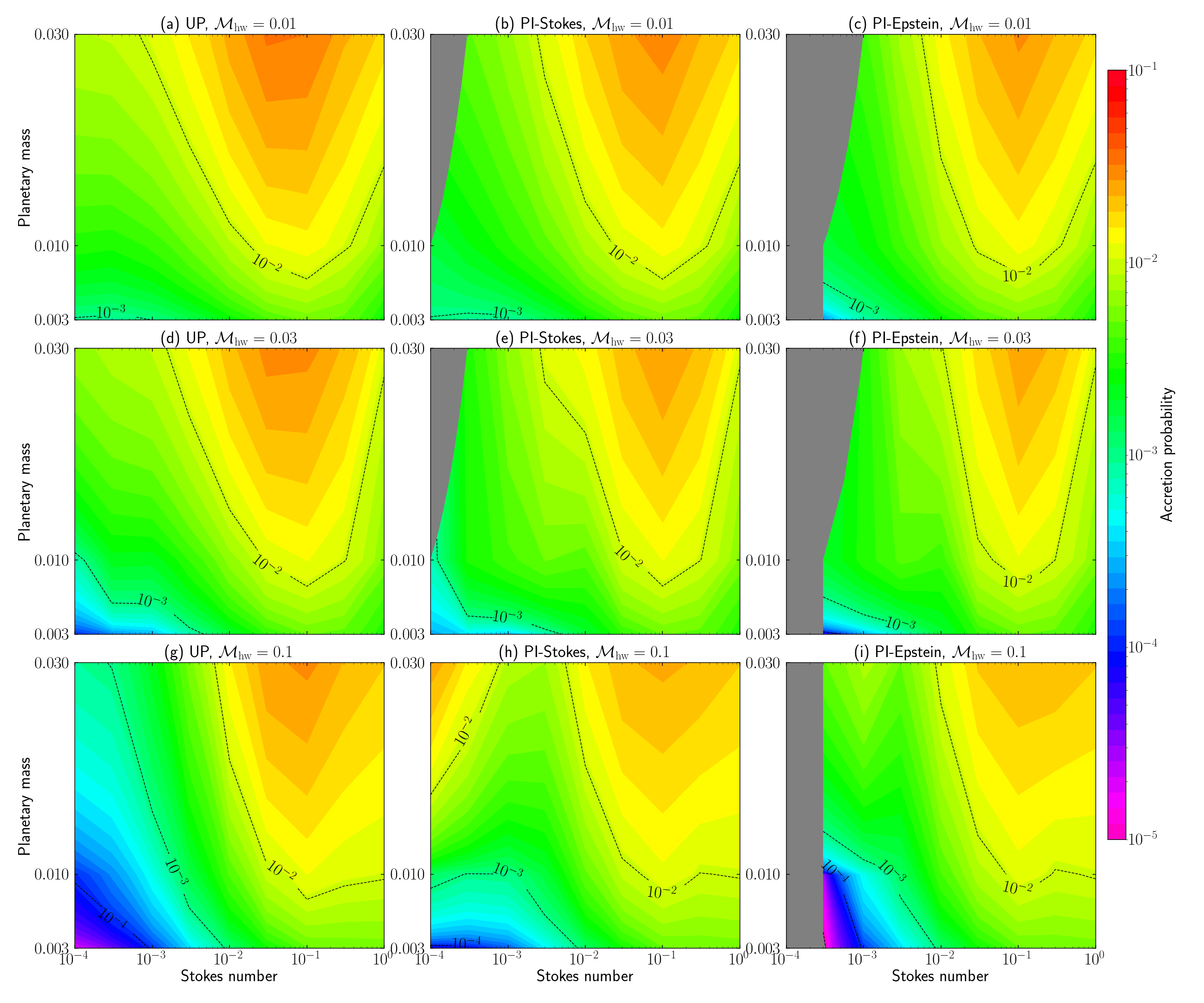}} 
 \caption{Accretion probability as a function of the planetary mass and the Stokes number for the Mach number $\mathcal{M}_{\rm hw}=0.01,\,0.03$ and $0.1$ (\textit{bottom to top}) in the \texttt{UP} case (\textit{left column}), the \texttt{PI-Stokes} case (\textit{middle column}), and the \texttt{PI-Epstein} case (\textit{right column}). The contours represent the accretion probabilities. We assumed $\alpha=10^{-3}$. The gray-shaded region is the region where $P_{\rm acc}=0$.}
\label{fig:Contour}
\end{figure*}
\fi

\subsection{Results overview}
The main subject of this study is to clarify the influence of the planet-induced gas flow on pebble accretion. In Sect. \ref{sec:result1}, we show the characteristic 3D structure of the planet-induced gas flow field obtained by 3D hydrodynamical simulations. In Sect. \ref{sec:result2}, we show the results of orbital calculations. Section \ref{sec:result3} shows the dependence of the accretion probability of pebbles on the planetary mass, the Stokes number, and the Mach number of the headwind of the gas.

In Sects. \ref{sec:result2} and \ref{sec:result3}, we classified the results into four categories according to the classification of the flow (Sects. \ref{sec:clasification} and \ref{sec:flow-transition}) and the accretion (Sect. \ref{sec:appendix1}) regimes as shown in \Figref{fig:summary}. The \texttt{UP} simulations were performed as control experiments in order to understand the influences of planet’s gravity by comparing the results with planet-induced flow (\texttt{PI}) ones, and thus \texttt{UP} cases are not categorized in any categories in \Figref{fig:summary}. The Keplerian and sub-Keplerian disk classification corresponds to different input parameter spaces ($\mathcal{M}_{\mathrm{hw}}=0$ and $\mathcal{M}_{\mathrm{hw}}\neq0$), and thus they do not correspond to four categories of the output results. Since the gas drag (Epstein and Stokes) regimes are determined independently of the planetary mass and the Mach number of the headwind, we do not use them in four categories in \Figref{fig:summary}. In other words, all four categories should have two sub-categories for gas drag regimes.

\subsection{Three-dimensional planet-induced gas flow}\label{sec:result1}
When $\mathcal{M}_{\rm hw}=0$, the planet-induced gas flow has a rotational symmetric structure with respect to the $z$-axis \citepalias[see Fig.2 of][]{Kuwahara:2020}. The nonzero headwind of the gas breaks the symmetry of the planet-induced gas flow \cite[]{Ormel:2015b,Kurokawa:2018}. Figure \ref{fig:gasflow-3D} shows the 3D flow structure around an embedded planet in an endmember case, $\mathcal{M}_{\rm hw}=0.1$. Gas flow shows three types of streamlines. (1) The planetary envelope is exposed to the headwind of the gas. Gas from the disk enters the Bondi sphere at low latitudes (inflow) and exits at high latitudes (outflow: the red lines of \Figref{fig:gasflow-3D}). This recycling flow passes the planet, tracing the surface of the isolated envelope whose size is $\lesssim0.5R_{\rm Bondi}$ \cite[]{Kurokawa:2018}.  The detailed structure of the recycling streamlines is shown in \Figref{fig:gasflow-3D-2}. (2) The horseshoe streamlines lie inside the planetary orbit (the green lines of \Figref{fig:gasflow-3D}). (3) The Keplerian shear flow extends inside the horseshoe flow and outside the planetary orbit (the blue lines of \Figref{fig:gasflow-3D}).  

\AK{\subsection{Classification of the planet-induced gas flow}\label{sec:clasification}}
We introduce classification of the planet-induced gas flow to clarify its influence on pebble accretion. Figure \ref{fig:gasflow-2D} shows how the structure at the midplane around an embedded planet depends upon the Mach number. When $\mathcal{M}_{\rm hw}=0.01$ and $0.03$, a slight rotational symmetry remains with respect to the $z$-axis (Figs. \ref{fig:gasflow-2D}a and b). The horseshoe streamlines still lie near the planetary orbit, which protects the planetary envelope from the headwind of the gas. The 3D structure of the planet-induced gas flow is similar to that found when $\mathcal{M}_{\rm hw}=0$, where the inflow occurs at high latitudes of the Bondi sphere and the outflow occurs at the midplane region of the disk \citepalias[see Fig.2 of][]{Kuwahara:2020}. We refer to such a case as the flow shear regime. All of the results shown in \citetalias{Kuwahara:2020} can be considered to be results in the flow shear regime. 
 
As the Mach number of the headwind of the gas increases, the horseshoe streamlines move to the negative direction in the $x$-axis. This is because the corotation radius for the gas moves toward the negative direction of the $x$-axis. When $\mathcal{M}_{\rm hw}=0.1$, the horseshoe streamlines lie inside the $x$-coordinate of the Bondi radius (\Figref{fig:gasflow-2D}c). The planetary envelope is exposed to the headwind. Within the Bondi radius, the azimuthal velocity of the gas is low (\Figref{fig:v-phi}). The envelope is pressure supported. The 3D structure of the planet-induced gas flow differs from that which can be seen in the flow shear regime. We refer to such a case as the flow headwind regime. Based on a series of hydrodynamical simulations, we classified the planet-induced gas flow into the flow shear and the flow headwind regimes, which are listed  in \Tabref{tab:1}. We discuss the transition from the flow shear to the flow headwind regime in Sect. \ref{sec:flow-transition}. 

\subsection{Orbital calculations}\label{sec:result2}
\subsubsection{Pebble accretion in 2D}
We first focus on the 2D limit of pebble accretion in the Stokes regime, that is all of the pebbles settled in the midplane of the disk and the Stokes number of pebbles does not depend on the gas density. Figure \ref{fig:peb-line1} shows the trajectories of pebbles at the midplane of the disk. We compared the results of \texttt{UP-m001-hw01} case and \texttt{PI-Stokes-m001-hw01} case. When the Stokes number is larger than ${\rm St}\geq10^{-1}$, the trajectories of pebbles and the width of the accretion window are similar (Figs. \ref{fig:peb-line1}c, d, g, and h). When the Stokes number is smaller than ${\rm St}\leq10^{-2}$, the trajectories of pebbles near the planetary orbit are deflected by the recycling flow (Figs. \ref{fig:peb-line1}a, b, e, and f). Since the planet chiefly perturbs the surrounding disk gas at a scale that is typically the smaller of the two when comparing the Bondi and Hill radii \cite[]{Kuwahara:2019}, the difference between  \texttt{UP-m001-hw01} and \texttt{PI-Stokes-m001-hw01} cases can be seen in the region close to the planet. In particular, the accretion window is wider in the planet-induced gas flow than in the unperturbed flow when ${\rm St}=10^{-3}$ (Figs. \ref{fig:peb-line1}a and e). This is in contrast to the conclusion of \citetalias{Kuwahara:2020}, where the width of the accretion window in the planet-induced gas flow in the flow shear regime becomes narrower than those in the unperturbed shear flow \citepalias[see Fig.3 of][]{Kuwahara:2020}. \AK{The difference is caused by the headwind of the gas.}

Figure \ref{fig:peb-line2} compares the results between \texttt{UP-m003-hw01} and \texttt{PI-Stokes-m003-hw01} cases. This figure shows the significant difference of the trajectories of pebbles with ${\rm St}=10^{-4}$. In the \texttt{PI-Stokes-m003-hw01} case, the recycling flow deflects the trajectories of pebbles. This deflection is not found in 2D simulations \cite[]{Ormel:2013}, but appears in the 3D ones. The pebbles are jammed outside the planetary orbit. Some of the pebbles from inside the planetary orbit move away from the planet along the horseshoe flow. In the flow shear regime, the horseshoe flow that lies near the planetary orbit reduces the width of the accretion window \citepalias{Kuwahara:2020}. The outflow occurs in the second and fourth quadrants of the $x$-$y$ plane \cite[]{Kuwahara:2019}. On the other hand, in the flow headwind regime, since the horseshoe flow shifts significantly to the negative direction in the $x$-axis, it does not suppress pebble accretion. The outflow occurs only in the fourth quadrant of the $x$-$y$ plane. The accretion of pebbles coming from the region where $y>0$ is not inhibited. Pebbles are susceptible to becoming entangled in the recycling flow. This leads to an increase in the time taken for pebbles to pass the Bondi radius of the planet (see Sect. \ref{sec:3.3.3}).

In the \texttt{PI-Epstein} case, the shape of trajectories of pebbles does not differ significantly from that in the \texttt{PI-Stokes} case. However, the width of the accretion window, the accretion cross section, and the accretion probability in the \texttt{PI-Epstein} case do not match those in the \texttt{PI-Stokes} case (see Sect. \ref{sec:result3}).
\subsubsection{Pebble accretion in 3D}\label{sec:3.3.2}
Next we focus on the 3D behavior of pebble accretion in the Stokes regime. Figure \ref{fig:peb-line3} shows the 3D trajectories of pebbles with ${\rm St}=10^{-4}$. This figure compares the results between \texttt{UP-m003-hw01} and \texttt{PI-Stokes-m003-hw01} cases. Since the gravity of the planet acting on the pebbles becomes weaker at high altitudes, pebbles do not accrete onto the planet in the \texttt{UP-m003-hw01} case (\Figref{fig:peb-line3}a). On the other hand, even if pebbles come from high altitudes ($z_{\rm s}\sim R_{\rm Bondi}$), they accrete onto the planet in the \texttt{PI-Stokes-m003-hw01} case (\Figref{fig:peb-line3}b). This is caused by the recycling flow in the vicinity of the planet. The typical scale of the recycling flow is the size of the Bondi radius (\Figref{fig:gasflow-3D-2}). In the flow shear regime, since the horseshoe flow has a vertical structure like a column, pebbles coming from high altitudes move away from the planet near the planetary orbit. In the flow headwind regime, the horseshoe flow does not inhibit pebble accretion. When pebbles that come from high altitudes reach the vicinity of the planet, a fraction of them that reside in $z\lesssim R_{\rm Bondi}$ become entangled in the recycling flow. This causes an increase in the time taken for pebbles to cross the Bondi radius (see Sect. \ref{sec:3.3.3}). Pebbles can accrete onto the planet even when they come from high altitudes. 
\subsubsection{Increase in the Bondi crossing time of pebbles}\label{sec:3.3.3}
Figures \ref{fig:v-peb1} and \ref{fig:v-peb2} show the trajectories of pebbles that are projected on the $x$-$y$ plane, and the relative velocity of pebbles to the planet as a function of the distance from the planet, $r$. We selected the pebbles passing near the Bondi region. In the \texttt{UP} cases, even when the pebbles pass near the Bondi sphere, their relative velocity does not change unless they  reach the region very close to the planet, $r\lesssim0.1R_{\rm Bondi}$ (Figs. \ref{fig:v-peb1}a and \ref{fig:v-peb2}a). In the \texttt{PI} cases, significant velocity fluctuation can be seen when the pebbles enter the Bondi sphere (Figs. \ref{fig:v-peb1}b and \ref{fig:v-peb2}b). We define the Bondi crossing time of pebbles as 
\begin{align}
    \tau_{\rm Bondi}=\frac{R_{\rm Bondi}}{v},\label{eq:t-bondi}
\end{align}
where $v$ is the relative velocity of pebbles to the planet. From Figs. \ref{fig:v-peb1}b and \ref{fig:v-peb2}b, the relative velocity of pebbles is reduced by an order of magnitude when they enter the Bondi sphere. This leads to an increase in the Bondi crossing time of pebbles by an order of magnitude. Just before accreting onto the planet, the relative velocity of pebbles to the planet reaches  terminal velocity in both \texttt{UP} and \texttt{PI} cases, which is determined by the force balance between the gas drag and the gravity of the planet acting on the pebble:
\begin{align}
    v_{\rm term}=\frac{m{\rm St}}{r^2}.\label{eq:terminal}
\end{align}

Within the Bondi sphere, the gas density increases significantly to maintain hydrostatic equilibrium. The velocity of the gas is reduced, and then a long-stagnant gas flow field is formed within the Bondi region (\Figref{fig:v-phi}). Once the pebbles enter the Bondi sphere and become entangled in the recycling flow, the strong gas drag force reduces their relative velocity.

In the flow headwind regime, the horseshoe flow shifts significantly to the negative direction in the $x$-axis and the outflow occurs only in the fourth quadrant in the $x$-$y$ plane. Pebbles coming from the region where $y>0$ and passing near the planetary orbit are susceptible to becoming entangled in the recycling flow. Thus, pebble accretion is enhanced in the planet-induced gas flow. 
\subsection{Accretion probability of pebbles}\label{sec:result3}
\subsubsection{Width of accretion window and accretion cross section}
Figure \ref{fig:w-acc} shows the changes of the width of the accretion window in the midplane region as a function of the Stokes number for different planetary masses and the Mach numbers. We first focus on the left column of \Figref{fig:w-acc}, where we compare the results between the \texttt{UP} case and \texttt{PI-Stokes} case. In common with all panels in the left column of \Figref{fig:w-acc}, the accretion window is wider in the planet-induced gas flow than in the unperturbed flow when the accretion and flow regime are both in the headwind regime (the filled circles in Figs. \ref{fig:w-acc}a, c, and e). This is one of the most important findings in our study. When the accretion occurs in the shear regime and the planet-induced gas flow is in the flow headwind regime (the filled squares in Figs. \ref{fig:w-acc}a, c, and e), the widths of the accretion window in the \texttt{UP} and \texttt{PI-Stokes} cases match each other. 

When the planet-induced gas flow is in the flow shear regime (the open circles and squares in Figs. \ref{fig:w-acc}a, c, and e), the trend can be explained by the conclusion of \citetalias{Kuwahara:2020}. In the flow shear regime, since the horseshoe flow lies near the planetary orbit, pebbles coming from the narrow region between the horseshoe and the shear regions can accrete onto the planet. This causes the reduction of the width of the accretion window. The width of the accretion window is not a simple increasing function of the Stokes number (e.g., Figs. \ref{fig:w-acc}a and b). As pebbles continue to drift inward while they are approaching the planet, the accretion of pebbles from $y<0$ does not occur at all in some cases, leading to the complicated dependence of accretion width on St (see \Figref{fig:Aacc-m001-hw003}).

When $m=0.01$ and $0.03$, the significant influence of the planet-induced gas flow can be seen for the pebbles with ${\rm St}\lesssim10^{-2}$. When $m=0.003$, the influence of the planet-induced gas flow is weak compared to the cases of $m\geq0.01$. The size of the perturbed region is determined by the gravity of the planet \cite[]{Kuwahara:2019}. Thus the influence of the planet-induced gas flow on pebble accretion becomes weak as the planetary mass decreases. 

Next we focus on the right column of \Figref{fig:w-acc}, where we compare the results between the \texttt{UP} case and \texttt{PI-Epstein} case. In contrast to the result shown in the left column of \Figref{fig:w-acc}, the width of the accretion window decreases in the planet-induced gas flow when the accretion and flow regime are both in the headwind regime (the filled circles in Figs. \ref{fig:w-acc}b, d, and f). Since the gas density is higher around the planet due to its gravity, the effective Stokes number decreases as the pebble approaches the planet. This causes significant reduction in the width of the accretion window, particularly for ${\rm St}\lesssim10^{-3}$. When the accretion occurs in the shear regime and the planet-induced gas flow is in the flow headwind regime (the filled squares in Figs. \ref{fig:w-acc}b, d, and f), or when the planet-induced gas flow is in the flow shear regime (the open circles and squares in Figs. \ref{fig:w-acc}b, d, and f), the results are similar to those in the \texttt{PI-Stokes} case.

 Figures \ref{fig:Aacc-m001-hw003} and \ref{fig:Aacc-m001-hw01} show the accretion windows of pebbles in the \texttt{UP-m001-hw003} case, \texttt{PI-Stokes-m001-hw003} case, \texttt{UP-m001-hw01} case, and in the \texttt{PI-Stokes-m001-hw01} case. We plotted all of the accretion windows in Figs. \ref{fig:Aacc-m001-hw003} and \ref{fig:Aacc-m001-hw01}.\footnote{In \citetalias{Kuwahara:2020}, we only plotted the accretion window in $x\geq0$ because the accretion windows have plane symmetry with respect to the $x=0$ plane \citepalias[see Fig.7 of][]{Kuwahara:2020}.}  In \Figref{fig:Aacc-m001-hw003}, there are one or two accretion windows. The accretion window has an asymmetric shape with respect to the $x=x_{\rm peb,\,cor}$ plane. This is caused by the radial drift of pebbles. For the range of the Stokes numbers considered here, the speed of the inward drift increases with the Stokes number and has a peak at ${\rm St}=10^{0}$ for a fixed Mach number (\Equref{eq:vx-int}). 
 
 We first focus on the top panel of \Figref{fig:Aacc-m001-hw003} (\texttt{UP-m001-hw003} case). Pebbles with ${\rm St}\geq10^{-2}$ coming from the region where $x<x_{\rm peb,\,cor}$ experience fast radial drift, and do not accrete onto the planet. When ${\rm St}=10^{-3}$, the accretion from the region where $x<x_{\rm peb,\,cor}$ can be seen due to the slow radial drift of pebbles (\Figref{fig:Aacc-m001-hw003}b). When ${\rm St}=10^{-4}$, radial drift of pebbles is limited, but the $x$-coordinate of the pebble corotation radius is larger than the maximum impact parameter of the accreted pebbles, $|x_{\rm peb,\,cor}|>b_{\rm hw}$. Thus, the accretion occurs only in the region where $x>x_{\rm peb,cor}$. The differences in the number of  accretion windows for the combination of ${\rm St},\,\mathcal{M}_{\rm hw}$ and $m$ lead to the complex behavior of the width of the accretion window (\Figref{fig:w-acc}). 
 
 In the bottom panels of \Figref{fig:Aacc-m001-hw003} (\texttt{PI-Stokes-m001-hw003} case), the shape of the accretion windows is similar to that in the \texttt{UP-m001-hw003} case when ${\rm St}\geq10^{-1}$. When ${\rm St}=10^{-2}$, the accretion occurs in the region where $x<x_{\rm peb,\,cor}$, which is not found in the \texttt{UP-m001-hw003} case. When ${\rm St}\leq10^{-3}$, the height of the accretion window in $x<x_{\rm peb,\,cor}$ is larger than that in $x>x_{\rm peb,\,cor}$. The 3D structure of the planet-induced gas flow is almost rotationally symmetric, but the polar inflow shifts  slightly to the negative direction in the $x$-axis as well as the horseshoe flow. This promotes the accretion of pebbles from high altitudes in the region where $x<x_{\rm peb,\,cor}$.

 Strong headwind causes fast radial drift. The accretion occurs only in the region where $x>x_{\rm peb.\,cor}$ in \Figref{fig:Aacc-m001-hw01}. In the flow headwind regime, pebbles can accrete onto the planet even if they come from high altitudes (\Figref{fig:peb-line3}). The width of the accretion window in the planet-induced gas flow is larger than that in the unperturbed flow (\Figref{fig:w-acc}). These findings can also be seen in \Figref{fig:Aacc-m001-hw01}. We found that the accretion window expands at the same location (Figs. \ref{fig:Aacc-m001-hw01}a and f). When the pebbles are well coupled to the gas (${\rm St}=10^{-4}$), the vertical scale of the accretion window is identical to the scale of the recycling flow ($z\sim R_{\rm Bondi}$). This is in contrast to the results in \citetalias{Kuwahara:2020}, where the outflow in the midplane region and the vertical structure of the horseshoe flow inhibits pebble accretion in the flow shear regime. 
 
 Figure \ref{fig:A-acc} shows the differences between the integrated accretion cross section in the \texttt{UP}, \texttt{PI-Stokes}, and \texttt{PI-Eptein} cases, and the dependence on the planetary mass and the Mach number. We first focus on the left column of \Figref{fig:A-acc}. As shown in \Figref{fig:Aacc-m001-hw01}, we confirmed that the accretion cross section is larger in the planet-induced gas flow than that in the unperturbed flow when the accretion and flow regime are both in the headwind regime (the filled circles in Figs. \ref{fig:A-acc}a, c, and e). This is an important finding, as is the increase in the width of the accretion window in the flow headwind regime. When the accretion occurs in the shear regime and the planet-induced gas flow is in the flow headwind regime (the filled squares in Figs. \ref{fig:A-acc}a, c, and e), the accretion cross sections in the \texttt{UP} and \texttt{PI-Stokes} cases match each other. When the planet-induced gas flow is in the flow shear regime (the open circles and squares in Figs. \ref{fig:A-acc}a, c, and e), the trend can be explained by the conclusion of \citetalias{Kuwahara:2020}. In the flow shear regime, since the horseshoe flow lies near the planetary orbit, pebbles coming from the narrow region between the horseshoe and the shear regions can accrete onto the planet. This causes a reduction of the accretion cross section in the planet-induced gas flow. Similarly to the width of the accretion window, the significant influence of the planet-induced gas flow can be seen for the pebbles with ${\rm St}\lesssim10^{-2}$ when $m=0.01$ and $0.03$, but it becomes weak when $m=0.003$.

Next we focus on the right column of \Figref{fig:A-acc}, where we compare the results between the \texttt{UP} case and \texttt{PI-Epstein} case. In contrast to the result shown in the left column of \Figref{fig:A-acc}, the accretion cross section is smaller in the planet-induced gas flow than in the unperturbed flow when the accretion and flow regime are both in the headwind regime, except when  $m=0.03$ (the filled circles in Figs. \ref{fig:A-acc}b and d). Since the gas density is higher around the planet due to its gravity, the effective Stokes number decreases as the pebble approaches the planet. This causes significant reduction in the width of the accretion window, in particular for ${\rm St}\lesssim10^{-3}$. When the accretion occurs in the shear regime and the planet-induced gas flow is in the flow headwind regime (the filled squares in Figs. \ref{fig:A-acc}b, d, and f), or when the planet-induced gas flow is in the flow shear regime (the open circles and squares in Figs. \ref{fig:A-acc}b, d, and f), the results are similar to those in the \texttt{PI-Stokes} case.
\subsubsection{Accretion probability}
Figure \ref{fig:Mdot-2D} shows the 2D accretion rate and accretion probability as a function of the Stokes number for the different planetary masses and the Mach numbers. The explanation in \Figref{fig:w-acc} can be applied to this figure. The location of the accretion window does not change in the planet-induced gas flow in the flow headwind regime. Figure \ref{fig:Mdot-2D} reflects the results of \Figref{fig:w-acc}. In the left column, where we compare the results between the \texttt{UP} case and \texttt{PI-Stokes} case, the 2D accretion probability is larger in the planet-induced gas flow than in the unperturbed flow when the accretion and flow regime are both in the headwind regime (the filled circles in Figs. \ref{fig:Mdot-2D}a, c, and e). This is because the width of the accretion window is larger in the planet-induced gas flow than that in the unpertubed flow.  We found that the enhancement of the 2D accretion probability in the \texttt{PI-Stokes} becomes more significant as the planetary mass increases.  We note that when $m=0.03$ and $\mathcal{M}_{\rm hw}=0.1$, the accretion probability of pebbles with ${\rm St}=10^{-4}$ approaches unity (\Figref{fig:Mdot-2D}e). Because we do not trace the pebble trajectory in the global disk, the accretion of slowly drifting pebbles might be double-counted in this limit \AK{\cite[]{Liu:2018}}. Thus we may overestimate the accretion probability.

When the accretion occurs in the shear regime and the planet-induced gas flow is in the flow headwind regime (the filled squares in Figs. \ref{fig:Mdot-2D}a, c, and e), the accretion probability in both the \texttt{UP} and \texttt{PI-Stokes} cases match each other because the width of the accretion window does not change in the planet-induced gas flow. When the planet-induced gas flow is in the flow shear regime (the open circles and squares in Figs. \ref{fig:A-acc}a, c, and e), the trend can be explained by the conclusion of \citetalias{Kuwahara:2020}. In the flow shear regime, the width of the accretion window is smaller in the planet-induced gas flow than that in the unperturbed flow. Thus the accretion probability is also smaller in the planet-induced gas flow than that in the unperturbed flow.\footnote{In contrast, when the planetary mass is large, $m\gtrsim0.1$, the reduction of the accretion cross section and the increase of relative velocity cancel each other out. Consequently, the accretion probability becomes comparable to that in the unperturbed flow \citepalias[see Sect. 3.4.2 of][]{Kuwahara:2020}.} 

In the right column of \Figref{fig:Mdot-2D}, the accretion probability is smaller in the planet-induced gas flow than in the unperturbed flow, when the accretion and flow regime are both in the headwind regime due to the significant reduction of the effective Stokes number in the vicinity of the planet, except when $m=0.03$ (the filled circles in Figs. \ref{fig:Mdot-2D}b and d). When the accretion occurs in the shear regime and the planet-induced gas flow is in the flow headwind regime (the filled squares in Figs. \ref{fig:Mdot-2D}b, d, and f), or when the planet-induced gas flow is in the flow shear regime (the open circles and squares in Figs. \ref{fig:Mdot-2D}b, d, and f), the results are similar to those in the \texttt{PI-Stokes} case. 

Figure \ref{fig:Mdot-3D-m001} shows the 3D accretion rate and accretion probability for a planet with $m=0.01$ as a function of the Stokes number for the different turbulent parameters and the Mach numbers. \AK{As seen in \Figref{fig:Mdot-3D-m001}, the 3D accretion probability is a decreasing function of $\alpha$. This is because the pebble scale height increases with $\alpha$ (\Equref{eq:pebble-scaleheight}).} We first focus on the left column of \Figref{fig:Mdot-3D-m001}, where we compare the results between the \texttt{UP} case and \texttt{PI-Stokes} case. When $\mathcal{M}_{\rm hw}=0.01$, the accretion probability in the \texttt{PI-Stokes} case matches or is slightly smaller than that in the \texttt{UP} case, albeit the accretion cross section is significantly smaller in the planet-induced gas flow than that in the unperturbed flow when ${\rm St}\lesssim10^{-3}$ (\Figref{fig:Mdot-3D-m001}a). This is because the reduction of the accretion cross section and the increase of relative velocity cancel each other out \citepalias{Kuwahara:2020}. When $\mathcal{M}_{\rm hw}=0.03$, the accretion probability in the \texttt{PI-Stokes} case has a double peak when $\alpha\lesssim10^{-4}$ (\Figref{fig:Mdot-3D-m001}c). The accretion window in the region where $x<x_{\rm peb,\,cor}$ disappear due to the radial drift of pebbles (\Figref{fig:Aacc-m001-hw003}), causing the local minimum of the accretion probability at ${\rm St}\sim10^{-2}$. 

When $\mathcal{M}_{\rm hw}=0.1$, similarly to the 2D case, the accretion probability in the \texttt{PI-Stokes} case is larger than that in the \texttt{UP} case (the filled circles in \Figref{fig:Mdot-3D-m001}e, where the accretion and flow regime are both in the headwind regime). When ${\rm St}=10^{-4}$, the achieved accretion probability in the \texttt{PI-Stokes} case is larger by an order of magnitude than that in the \texttt{UP} case. When the accretion occurs in the shear regime and the planet-induced gas flow is in the flow headwind regime (the filled squares in \Figref{fig:Mdot-3D-m001}e), the accretion probability in both \texttt{UP} and \texttt{PI-Stokes} cases match each other. 

In the right column of \Figref{fig:Mdot-3D-m001}, the aforementioned features can be seen in common when ${\rm St}\gtrsim10^{-3}$. Only when ${\rm St}\lesssim10^{-3}$, the achieved accretion probability in the \texttt{PI-Epstein} case is smaller than that in the \texttt{UP} case (Figs. \ref{fig:Mdot-3D-m001}b, d, and f). The 3D accretion probabilities for the different planetary masses are shown in Figs. \ref{fig:Mdot-3D-m0003} and \ref{fig:Mdot-3D-m003}. These figures also show a trend to that in \Figref{fig:Mdot-3D-m001}. We found that the enhancement of the 3D accretion probability in the \texttt{PI-Stokes} case becomes significant as the planetary mass increases. 

Figure \ref{fig:Contour} shows the accretion probability as a function of both the planetary mass and the Stokes number for the various Mach numbers, $\mathcal{M}_{\rm hw}$. We fixed turbulence strength, $\alpha=10^{-3}$. When $\mathcal{M}_{\rm hw}=0.01$ (Figs. \ref{fig:Contour}a--c), a peak of accretion probability appears in the upper right region, where the planetary mass and the Stokes number are large. In the planet-induced gas flow, the accretion of small pebbles (${\rm St}\lesssim10^{-3}$) is suppressed. When $\mathcal{M}_{\rm hw}=0.03$ (Figs. \ref{fig:Contour}d--f), the region where $P_{\rm acc}\lesssim10^{-3}$ (lower left) expands compared to the results when $\mathcal{M}_{\rm hw}=0.01$. As the Mach number increases, the approach velocity of pebbles increases. This reduces encounter time in which pebbles experience the gravitational pull of the planet. Thus, when $\mathcal{M}_{\rm hw}=0.1$, the region where $P_{\rm acc}\lesssim10^{-3}$ is wider than the case when $\mathcal{M}_{\rm hw}\leq0.03$. We note that, when $\mathcal{M}_{\rm hw}=0.1$, two peaks of accretion probability appear in \Figref{fig:Contour}h. A peak that lies in the upper left region in \Figref{fig:Contour}h shows the enhancement of pebble accretion in the flow headwind regime. When the Stokes gas drag law is adopted, and the accretion and the flow regime are both in the headwind regime, the accretion probability of pebbles in the planet-induced gas flow is larger than that in the unperturbed flow.

\iffigure
 \begin{figure}[htbp]
 \resizebox{\hsize}{!}
 {\includegraphics{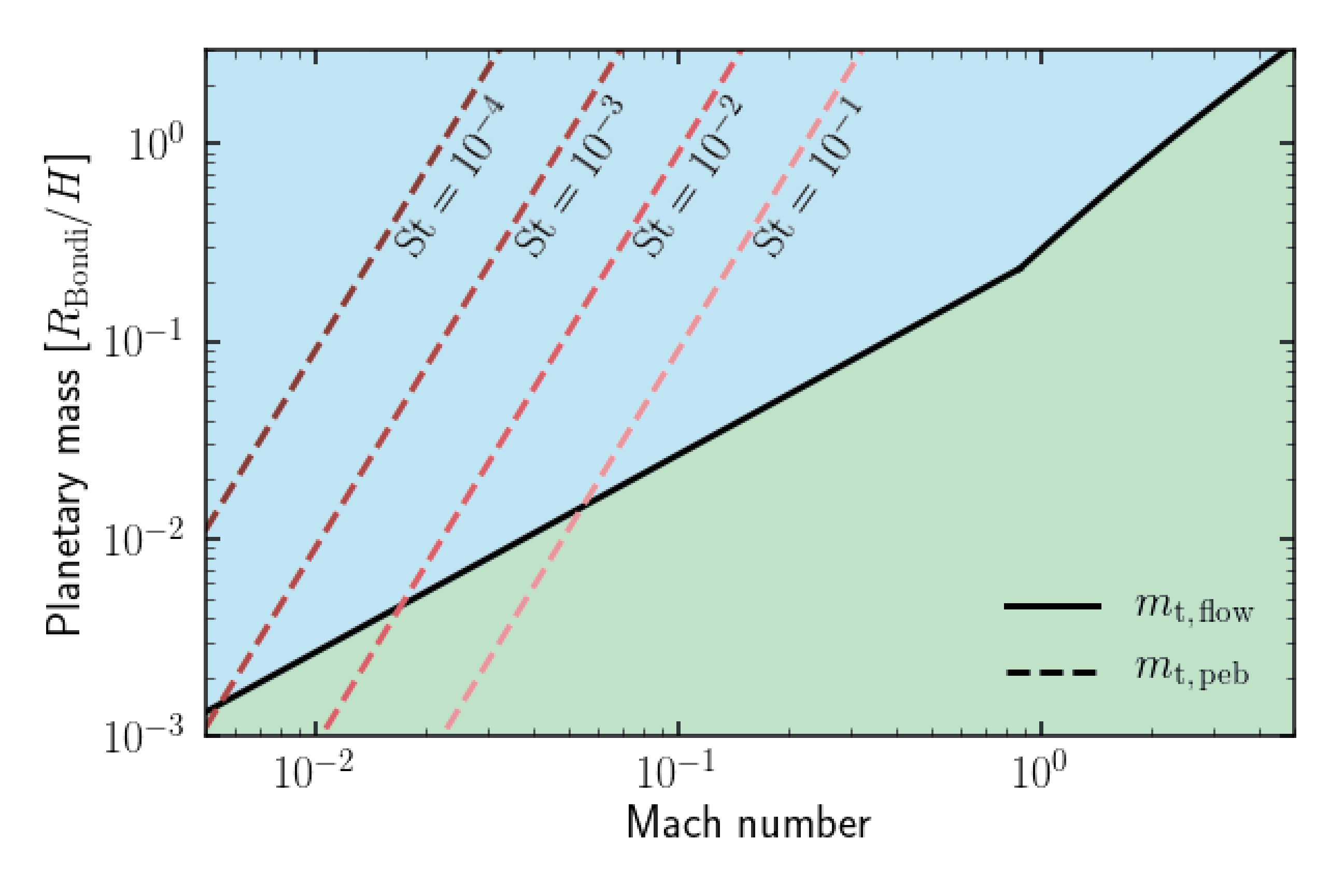}} 
 \caption{Flow transition mass as a function of the Mach number of the headwind of the gas (black solid line). The dashed lines correspond to the pebble transition mass with different Stokes numbers (\Equref{eq:pebble-transition}). \AK{The blue and green regions correspond to the flow shear and the flow headwind regime, respectively. Since our study do not handle gap formation (high-mass planets; Sect. \ref{sec:high-mass}), the range of the vertical axis was set to $m\leq3$.}}
\label{fig:flow-transition}
\end{figure}
\fi

\iffigure
 \begin{figure*}[htbp]
 \resizebox{\hsize}{!}
 {\includegraphics{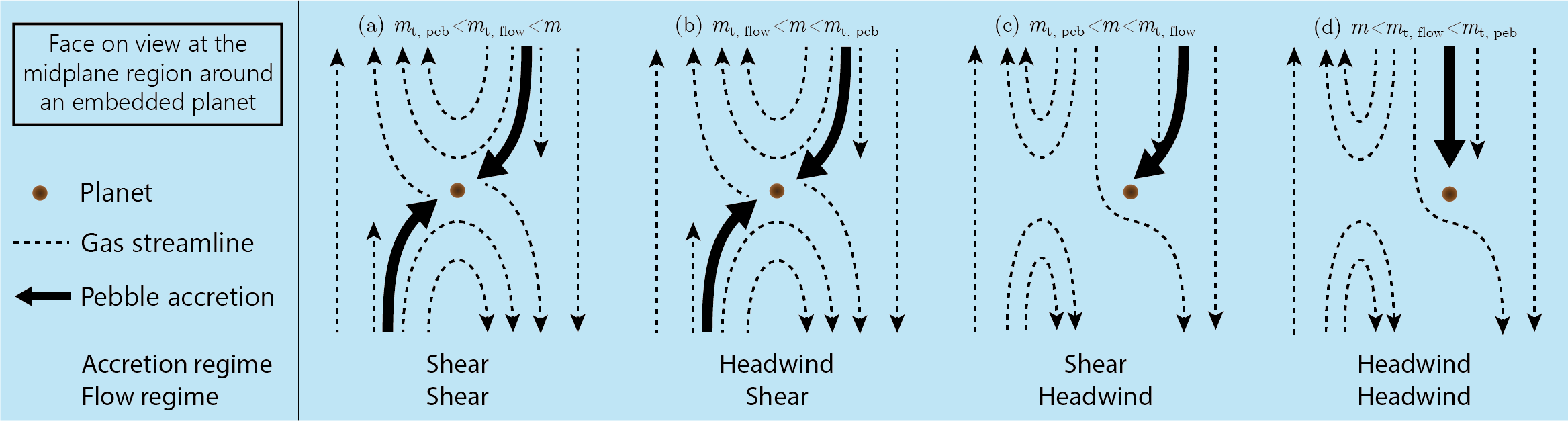}} 
 \caption{Schematic illustration of the flow structure and the trajectories of accreted pebbles at the midplane region. We classify the results obtained in both \citetalias{Kuwahara:2020} and this study into four categories based on the relation between $m,\,m_{\rm t,\,flow}$ and $m_{\rm t,\,peb}$. We note that the gas streamlines and the trajectories of accreted pebbles are rough outlines, and may differ slightly from the actual ones.}
\label{fig:schematic}
\end{figure*}
\fi

\iffigure
 \begin{figure}[htbp]
 \resizebox{\hsize}{!}
 {\includegraphics{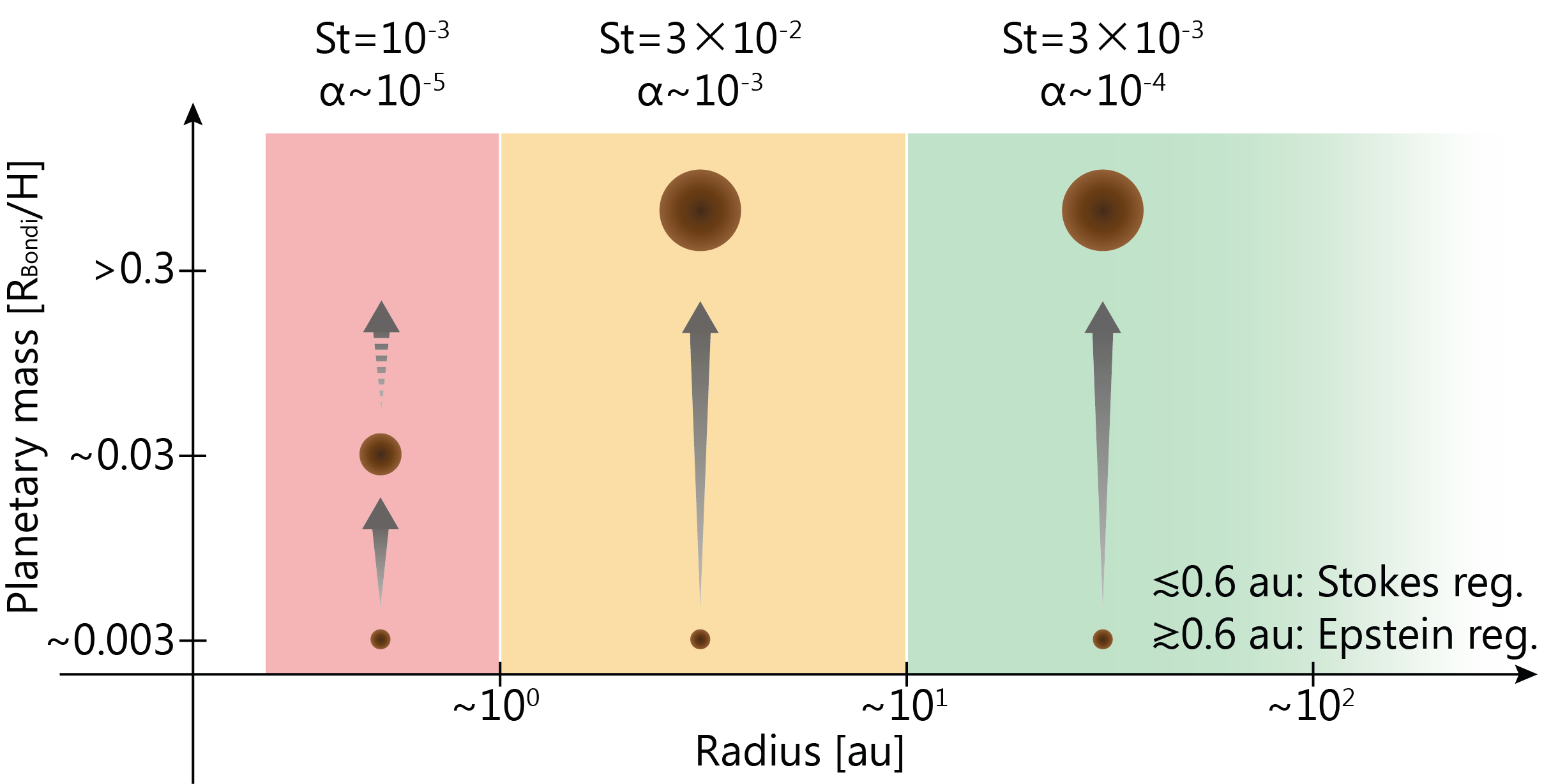}} 
 \caption{Schematic illustration of the growth of protoplanets. The brown filled circles denote the protoplanets. The assumed Stokes numbers and the turbulence strengths are shown. The transition from the Stokes to the Epstein regime occurs at $\sim0.6$ au in our parameter set. In the early phase of planetary growth, the planet-induced gas flow does not inhibit pebble accretion for a range of the Stokes numbers considered here. When the planetary mass reaches $m\sim0.03$ (the planet-induced flow isolation mass, $m_{\rm PI,\,iso}$)}, pebble accretion begins to be suppressed only in the inner region of the disk. The subsequent growth of the protoplanets in the inner region of the disk is highly suppressed \citepalias[the dashed arrow;][]{Kuwahara:2020}
\label{fig:schematic2}
\end{figure}
\fi

\section{Discussion}\label{sec:discussion}
\subsection{Flow transition mass}\label{sec:flow-transition}
We describe the transition from the flow shear to the flow headwind regime in Sect. \ref{sec:clasification}. In this section, we derive an analytical estimation that distinguishes these hydrodynamical regimes. A planet has an isolated envelope whose size is $\sim0.5R_{\rm Bondi}$ \cite[]{Kurokawa:2018}. The horseshoe streamlines are formed at the corotation radius for the gas. Here we define the flow transition mass, which describes the transition from the flow shear to flow headwind regime. It is given by the solution of the following equation: 
\begin{align}
    x_{\rm g,\,cor}+w_{\rm HS}=-\frac{m_{\rm t,\,flow}}{2}.\label{eq:HS-envelope}
\end{align}
The left-hand side of \Equref{eq:HS-envelope} corresponds to the maximum $x$-coordinates of the horseshoe region (the right edge). The right-hand side of \Equref{eq:HS-envelope} corresponds to the minimum $x$-coordinate of an isolated envelope (left edge). The width of the horseshoe region can be described by \cite[]{Masset:2016}
\begin{align}
    w_{\rm HS}=1.05\gamma^{-1/4}\sqrt{m}.\label{eq:w-HS1}
\end{align}
As discussed in Sect. 4.2.1 in \citetalias{Kuwahara:2020}, \Equref{eq:w-HS1} agrees with the half width of the horseshoe region of our hydrodynamical simulation when $m\gtrsim0.1$, but is otherwise an overestimation. From our series of hydrodynamical simulations, we found that the half width of the horseshoe region for the range of  planetary masses considered in this study can be described by  
\begin{align}
    w_{\rm HS}\simeq 2m.\label{eq:w-HS2}
\end{align}
We found that the width of the horseshoe region decreases slightly as the Mach number increases, but it does not decrease by an order. Considering the detailed scaling with $\mathcal{M}_{\rm hw}$ is beyond the scope of this study. We assume that the width of the horseshoe region for the range of the planetary masses considered in this study is always given by \Equref{eq:w-HS2}. From Eqs. (\ref{eq:HS-envelope}), (\ref{eq:w-HS1}) and (\ref{eq:w-HS2}), the flow transition mass can be described by
\begin{empheq}[left={m_{\rm t,\,flow}=\empheqlbrace}]{alignat=2}
&\left(-1.05\gamma^{-1/4}+\sqrt{1.1025\gamma^{-1/2}+\frac{4}{3}\mathcal{M}_{\rm hw}}\right)^{2}, \label{eq:m-flow-1} \\
&\hspace{120pt} \text{(for $m\gtrsim0.1$)},\nonumber \\ 
&\frac{4}{15}\mathcal{M}_{\rm hw},\hspace{83pt} \text{(for $m\lesssim0.1$)},\label{eq:m-flow-2}
\end{empheq}
where we only take the positive root in \Equref{eq:m-flow-1}.  We plotted the larger of Eqs. (\ref{eq:m-flow-1}) and (\ref{eq:m-flow-2}) in \Figref{fig:flow-transition}. We note that we do not consider the reduction of the width of the horseshoe region due to the strong headwind of the gas. Thus we may underestimate the flow transition mass, in particular when $\mathcal{M}_{\rm hw}\gtrsim0.1$. In the MMSN model, the Mach number of the headwind has an order of $\sim0.1$ even in the outer region of the disk ($\sim100$ au). Thus \Equref{eq:m-flow-2} can be applied to a wide range of our disk model. From \Equref{eq:m-flow-2}, the dimensional flow transition mass in the MMSN model can be described by 
\begin{align}
    M_{\rm t,\,flow}=0.16\left(\frac{a}{1\,\text{au}}\right)\,M_{\oplus},\qquad (\text{for $\mathcal{M}_{\rm hw}\lesssim0.9$}).
\end{align}
From \Figref{fig:flow-transition}, we can divide our results into four categories:
\begin{enumerate}
    \item When $m_{\rm t,\,peb}<m_{\rm t,\,flow}<m$, where $m_{\rm t,\,peb}$ is the transition mass for pebble accretion (\Equref{eq:pebble-transition}), the accretion and flow regime are both in the shear regime (\Figref{fig:schematic}a). As shown in \citetalias{Kuwahara:2020}, the accretion probability of pebbles in the \texttt{PI-Stokes} case matches when ${\rm St}\gtrsim3\times10^{-3}$--$10^{-2}$, or is smaller than that in the \texttt{UP} case when ${\rm St}$ falls below the preceding value in 2D. When $m=0.003$, the influence of the planet-induced gas flow is too weak to affect pebble accretion for the range of the Stokes number considered in this study. In 3D, the accretion probability in \texttt{PI-Stokes} case matches or is slightly larger (smaller) than that in the \texttt{UP} case when $m\geq0.1$ ($m\leq0.03$) for ${\rm St}\lesssim10^{-3}$--$10^{-2}$. The width of the horseshoe region increases as the planetary mass increases. The reduction of the accretion cross section and the increase of relative velocity cancel each other out when $m\geq0.1$. In the \texttt{PI-Epstein} case, since the reduction of the accretion window becomes more significant than those in the \texttt{PI-Stokes} case, the increase of the relative velocity does not fully offset its reduction. Therefore, the accretion probability in the \texttt{PI-Epstein} case tends to be smaller than that in the \texttt{UP} case, both in 2D and in 3D.
    
    \item When $m_{\rm t,\,flow}<m<m_{\rm t,\,peb}$, the accretion occurs in the headwind regime, but the planet-induced gas flow is in the shear regime (\Figref{fig:schematic}b). As in the case above, St satisfies the following relation in our parameter space: ${\rm St}\lesssim10^{-3}$. The pebbles tend to follow the gas streamlines. The conclusion from \citetalias{Kuwahara:2020} can be applied. The accretion window decreases in the planet-induced gas flow. The reduction of the accretion window and the increase of relative velocity due to the horseshoe flow cancel each other out.
    
    \item When $m_{\rm t,\,peb}<m<m_{\rm t,\,flow}$, the accretion occurs in the shear regime, but the planet-induced gas flow is in the flow headwind regime (\Figref{fig:schematic}c). As in the case above, St satisfies the following relation in our parameter space: ${\rm St}\gtrsim10^{-3}$. Pebbles are less affected by the gas flow. Pebbles coming from the region where $x<x_{\rm peb,\,cor}$ experience fast radial drift, and do not accrete onto the planet. The accretion probability in both \texttt{PI-Stokes} and \texttt{PI-Epstein} cases match that in the \texttt{UP} case.
    
    \item When $m<m_{\rm t,\,flow}<m_{\rm t,\,peb}$, the accretion and flow regime are both in the headwind regime (\Figref{fig:schematic}d). Pebbles coming from the region where $x<x_{\rm peb,\,cor}$ experience fast radial drift, and do not accrete onto the planet. The accretion probability is larger in the \texttt{PI-Stokes} case than that in the \texttt{UP} case. In the \texttt{PI-Epstein} case, the accretion probability is smaller than that in the \texttt{UP} case. We found a significant enhancement of pebble accretion when $m\geq0.01$ both in 2D and in 3D. The achieved accretion probability for the pebbles with ${\rm St}=10^{-4}$ in the \texttt{PI-Stokes} case is larger by an order of magnitude than that in the \texttt{UP} case.
\end{enumerate}

\subsection{Comparison to previous studies}
Assuming 2D and inviscid fluid, \cite{Ormel:2013} derived the steady state solution of 2D flow around an embedded planet. The author calculated the trajectories of small particles using the derived flow pattern. \cite{Ormel:2013} showed the trajectories of pebbles with ${\rm St}=10^{-4},\,10^{-3}$, and $10^{-2}$ around the planet with $m=0.01$ (Fig.12 of \cite{Ormel:2013}). Two cases are shown: $\mathcal{M}_{\rm hw}=0$ and $\mathcal{M}_{\rm hw}=0.05$. The latter case satisfies $m<m_{\rm t,\,flow}<m_{\rm t,\,peb}$ for ${\rm St}\leq10^{-1}$, where the accretion and flow regime are both in the headwind regime. Thus, the accretion of pebbles is expected to be promoted based on our results. However, accretion of small dust particles is suppressed in \cite{Ormel:2013}. A plausible reason is that the size of the atmosphere is different in 2D and 3D. \cite{Ormel:2013} found that the averaged size of the atmosphere in 2D is $\sim R_{\rm Bondi}$ when $m=0.01$ and $\mathcal{M}_{\rm hw}=0.05$. The small dust particles follow the gas streamlines outside the Bondi radius, and they passed the planet without breaking through the atmosphere. In our study, the size of an isolated envelope is $\lesssim0.5R_{\rm Bondi}$ \cite[]{Kurokawa:2018}. In 3D, the small size of an isolated envelope allows pebbles to approach close to the planet. The extension of the Bondi crossing time of pebbles due to the recycling flow further promotes pebble accretion.

\cite{Rosenthal:2018a} introduced flow isolation mass, $m_{\rm flow}^{\rm R18}$, as the solution of $R_{\rm Bondi}=R_{\rm Hill}$. In our dimensionless unit, the flow isolation mass is described by $m_{\rm flow}^{\rm R18}=0.58$. Based on the analytical argument without the influence of the planet-induced gas flow, these latter authors found that pebble accretion for all pebble sizes is inhibited when the planetary mass exceeds the flow isolation mass, $m>m_{\rm flow}^{\rm R18}$. When $m>m_{\rm flow}^{\rm R18}$, we found that $m>m_{\rm t,\,flow}$ for a wide range of the disk where $\mathcal{M}_{\rm hw}\lesssim1$ (\Figref{fig:flow-transition}). When the planetary mass reaches the flow isolation mass \cite[]{Rosenthal:2018a}, the planet-induced gas flow is in the flow shear regime. In the flow shear regime, we found that the accretion of pebbles with ${\rm St}\lesssim10^{-3}$ is suppressed significantly when we assumed the Epstein gas drag regime, but the accretion probability of pebbles with ${\rm St}\gtrsim10^{-3}$ in the planet-induced gas flow is comparable to that in the unperturbed flow \citepalias{Kuwahara:2020}. The difference is likely due to the smaller size of the bound atmosphere and complicated recycling flow patterns, both of which were not taken into account in \cite{Rosenthal:2018a}.

Moreover, \cite{Kuwahara:2019} found that the suppression of pebble accretion by the gas flow would not be expected when ${\rm St}\gtrsim0.4$, even for the higher-mass planets ($m>0.3$). As the planetary mass increases, the speed of the outflow at the midplane region increases \cite[]{Kuwahara:2019}. Our simulations are performed for the planetary mass with at most $m=0.3$ \citepalias{Kuwahara:2020}, the suppression of pebble accretion due to the gas flow might be prominent for the larger Stokes number when we assume $m>0.3$. However, comparing the outflow speed to the terminal velocity of pebbles, \cite{Kuwahara:2019} found that the suppression of pebble accretion due to the midplane outflow is limited to ${\rm St}\lesssim0.4$ \cite[see Fig. 9 of][]{Kuwahara:2019}. 

\subsection{Implications for the growth of protoplanets}
In \citetalias{Kuwahara:2020}, assuming the distribution of the turbulence strength and the size of the solid materials, we proposed a formation scenario of planetary systems to explain the distribution of exoplanets (the dominance of super-Earths at $<1$ au \cite[]{Fressin:2013,Weiss:2014} and a possible peak in the occurrence of gas giants at $\sim2$--$3$ au \cite[]{Johnson:2010,Fernandes:2019}, as well as the architecture of the Solar System). We divided the disk into three sections according to previous studies and assumed turbulence strength in each section as: $\alpha\sim10^{-5}$ ($\lesssim1$ au), $\alpha\sim10^{-3}$ ($\sim1$--$10$ au), and $\alpha\sim10^{-4}$ ($\sim10$ au) \cite[]{Malygin:2017,Lyra:2019}. Given the size distribution of the solid materials in a disk \cite[]{Okuzumi:2019}, we assumed that the pebbles have ${\rm St}\sim10^{-3}$ ($\lesssim1$ au), ${\rm St}\sim3\times10^{-2}$ ($\sim1$--$10$ au), and ${\rm St}\sim3\times10^{-3}$ ($\sim10$ au; \Figref{fig:schematic2}). In \citetalias{Kuwahara:2020}, we considered the growth of the protoplanet with $m\sim0.03$. The accretion and flow regime were both in the shear regime. In other words, we focused on the late stage of planet formation. Here we adopt the same assumption for the distribution of the turbulence strength and the size of the solid materials to be consistent, but consider the growth of the protoplanets with $m\sim0.003$. Thus, we now consider an earlier phase of planet formation compared to that in \citetalias{Kuwahara:2020}.

\subsubsection{Pebble accretion in smooth disks}\label{sec:4.3.1}
We first consider the growth of the protoplanets in a smooth disk. We do not consider any substructures in a disk (e.g., the gaps and rings). The Mach number in the MMSN model is given by $\mathcal{M}_{\rm hw}\simeq0.05\,(a/1\,{\rm au})^{1/4}$. The planet-induced gas flow around the planet with $m=0.003$ is in the flow headwind regime for a wide range of the disk ($a\gtrsim0.1$ au). In contrast to \citetalias{Kuwahara:2020}, where the achieved accretion probability in the planet-induced gas flow was very low ($P_{\rm acc}\sim3\times10^{-4}$ ) compared to that in the unperturbed flow ($P_{\rm acc}\sim7\times10^{-2}$) for $m=0.03$ at the inner region of the disk ($<1$ au), we would expect that the accretion probability in the planet-induced gas flow would be almost identical to that in the unperturbed flow across the entire region of the disk. From \Figref{fig:Contour}, the planet-induced gas flow has little effect on the accretion probability for the range of the Stokes number assumed here (${\rm St}\geq10^{-3}$). Thus, in the early phase of planet formation, the growth rate of the protoplanets can be estimated by the analytical arguments which is developed in the unperturbed flow \cite[e.g.,][]{Ormel:2017-pebble,Liu:2018,Ormel:2018}. Only in exceptional cases, where $m<m_{\rm t,\,flow}<m_{\rm t,\,peb}$, ${\rm St}\lesssim10^{^-3}$, and the Stokes drag law is adopted, the growth of the protoplanets would be accelerated (\Figref{fig:Contour}h). When the planetary mass reaches $m\gtrsim0.03$ ($M_{\rm pl}\gtrsim0.36\,M_{\oplus}$ at $1$ au), the accretion of small pebbles (${\rm St}\lesssim10^{-3}$) in the planet-induced gas flow begins to be suppressed in the region where $\mathcal{M}_{\rm hw}<0.1$ (Figs. \ref{fig:Contour}c and f).

\subsubsection{Pebble accretion at the pressure bump}\label{sec:4.3.2}
Recent observations show the gas and ring structures in a disk \cite[e.g.,][]{ALMA:2015,Pinte:2015,Andrews:2016,Isella:2016,Cieza:2017,Fedele:2018,Andrews:2018,Long:2018,Dullemond:2018,Van:2019,Long:2020}. Several mechanisms have been proposed to explain the ring structure in a disk: the dust accumulation at the edge of the dead-zone \cite[]{Flock:2015}, dust growth at snow lines \cite[]{Zhang:2015,Okuzumi:2016}, secular gravitational instability \cite[]{Takahashi:2014,Takahashi:2016,Tominaga:2018,Tominaga:2019}, or gap opening due to the presence of planets \cite[]{Kanagawa:2015,Kanagawa:2016,Dong:2015,Dipierro:2015}. The possible origin of the ring is still debated. When the ring structure is related to the dust getting trapped in radial pressure bumps \cite[]{Dullemond:2018}, where $\mathcal{M}_{\rm hw}=0$, the contribution of the headwind of the gas vanishes. Even in the outer region of the disk, the accretion and flow regime are both in the shear regime at the pressure bump.

Figures \ref{fig:Contour}b and c show the accretion probability for $\mathcal{M}_{\rm hw}=0.01$, but correspond to the case where the accretion and flow regime are both in the shear regime. Thus, we can estimate the accretion probability for the planet with $m=0.003$ at the pressure bump from Figs. \ref{fig:Contour}b and c. From Figs. \ref{fig:Contour}b and c, the accretion probability in the planet-induced gas flow for the planet with $m=0.003$ is identical to that in the unperturbed flow for a range of the Stokes number considered here, ${\rm St}\geq10^{-3}$. Same as in the smooth disks, in the early phase of planet formation, the growth rate of the protoplanets can be estimated by the analytical arguments which is developed in the unperturbed flow \cite[e.g.,][]{Ormel:2017-pebble,Liu:2018,Ormel:2018}. At the pressure bump, the enhancement of pebble accretion due to the planet-induced gas flow would never occur. \AK{Nevertheless, an increase in the dust-to-gas ratio at the pressure bump would lead to an increase in the accretion rate of pebbles.} When the planetary mass reaches $m\gtrsim0.03$ ($M_{\rm pl}\gtrsim\AK{0.36\,M_{\oplus}}$ at $1$ au), the suppression of pebble accretion for ${\rm St}\lesssim10^{-3}$ in the planet-induced gas flow becomes prominent (Figs. \ref{fig:Contour}b and c, see also Figs. 10 and 11 in \citetalias{Kuwahara:2020}).

\subsubsection{Pebble accretion for high-mass planets}\label{sec:high-mass}
Out study focused on low-mass planets (\Figref{fig:summary}). High-mass planets shape global pressure gradient, and an induced pressure maximum inhibits pebble accretion \cite[]{Lambrechts:2014,Bitsch:2018}. Planets with $m>3$ form a gap in a disk \cite[]{Lin:1993}. Pebble isolation initiates at a lower planetary mass in a disk with smaller turbulent viscosity \cite[]{Bitsch:2018}. \cite{Fung:2018} reported that a planet with $m\simeq0.4$ ($6.6\,M_{\oplus}$ at $\sim1.5$ au) can form a pressure bump in an inviscid disk.

\subsubsection{Planet-induced flow isolation mass}
The dimensionless planetary masses of $0.03$ can be regarded as a type of isolation mass, in particular when ${\rm St}\lesssim10^{-3}$ and $\mathcal{M}_{\rm hw}\lesssim0.1$. In such a specific case, pebble accretion would be expected to halt before the mass of the protoplanets reaches pebble isolation mass:
\begin{align}
    M_{\rm iso}^{\rm L14}\simeq20\,\left(\frac{a}{5\,{\rm au}}\right)^{3/4}\,M_{\oplus},
\end{align}
\cite[]{Lambrechts:2014}. A subsequent study derived a detailed pebble isolation mass:
\begin{align}
    M_{\rm iso}\simeq25\left(\frac{H/r}{0.05}\right)^3\left[0.34\left(\frac{3}{\log\alpha}\right)^4+0.66\right]\,M_{\oplus},\label{eq:Miso}
\end{align}
\cite[]{Bitsch:2018}. Equation (\ref{eq:Miso}) gives
\begin{align}
    M_{\rm iso}\simeq7.2\left(\frac{a}{1\,{\rm au}}\right)^{3/4}\,M_{\oplus},\label{eq:Miso2}
\end{align}
when we assume $H/r\simeq0.033(a/1\,{\rm au})^{1/4}$ and $\alpha=10^{-3}$. In our dimensionless unit, \Equref{eq:Miso2} can be described by
\begin{align}
    m_{\rm iso}\simeq0.6.\label{eq:Miso3}
\end{align}
When we refer $m=0.03$ as planet-induced flow isolation mass, $m_{\rm PI,\,iso}$, and then we have
\begin{align}
    m_{\rm PI,\,iso}=0.05m_{\rm iso}.
\end{align}

\subsection{Implications for the origins of the architecture of a planetary system}
In \citetalias{Kuwahara:2020}, tracking the growth of the planet with $m=0.03$, we introduced a formation scenario of planetary systems, where pebble accretion is suppressed only in the inner region of the disk ($\lesssim1$ au). As discussed in Sects. \ref{sec:4.3.1} and \ref{sec:4.3.2}, the planet-induced gas flow does not affect the accretion probability of pebbles in the earlier phase of planet formation compared to the situation that was considered in \citetalias{Kuwahara:2020} (\Figref{fig:schematic2}).

When we focus on the region where $<100$ au ($\mathcal{M}_{\rm hw}\lesssim0.1$), and the planetary mass exceeds $m>0.03$, the planet-induced gas flow is always in the flow shear regime (\Figref{fig:flow-transition}). Based on the discussion in \citetalias{Kuwahara:2020}, we propose a possible scenario for the distribution of the exoplanets again:
\begin{enumerate}
    \item The rocky terrestrial planets or super-Earths are formed in the inner region of the disk, $\lesssim1$ au, where $\alpha$ and St are small. The protoplanets accrete pebbles without the influence of planet-induced gas flow, and reach planet-induced flow isolation mass, $m_{\rm PI,\,iso}$ ($\sim0.36\,M_{\oplus}$ at $1$ au). When the mass of the protoplanets reaches $m\gtrsim m_{\rm PI,\,iso}$, the suppression of pebble accretion due to the planet-induced gas flow is prominent \citepalias{Kuwahara:2020}. The subsequent growth of the protoplanets is highly suppressed. Within the growth track, the small rocky terrestrial planets may experience inward migraiton. Inside $\lesssim0.6$ au, the gas drag law switches to the Stokes regime in our parameter set. This leads to an increase in the accretion probability of pebbles \citepalias{Kuwahara:2020}. The small rocky protoplanets may also experience giant impacts. These events lead to the formation of super-Earths.
    \item The gas giants are formed in the middle region of the disk, $\sim1$--$10$ au, where $\alpha$ and St are larger than those in the inner region. Since the Stokes number is large, the planet-induced gas flow does not inhibit the growth of the protoplanets even though the mass of the protoplanets reaches $m\gtrsim m_{\rm PI,\,iso}$ \citepalias{Kuwahara:2020}. Thus, we expect that the planets might exceed the critical core mass within the typical lifetime of the disk. \AK{The final mass of the protoplanets is set by the pebble isolation due to formation of pressure bump \cite[]{Lambrechts:2014,Bitsch:2018}.}
    \item The ice giants are formed in the outer region, where $\alpha$ and St are smaller than those in the middle region. In the same way as in the middle region, planet-induced gas flow does not inhibit the growth of the protoplanets until their mass reaches $m\sim m_{\rm PI,\,iso}$. When $m\gtrsim m_{\rm PI,\,iso}$, the accretion probability in the planet-induced gas flow is reduced by a factor compared to that in the unperturbed flow \citepalias{Kuwahara:2020}. With the accretion of a moderate amount of icy pebbles, the planets would eventually become ice giants without evolving into gas giants.
\end{enumerate}
Our scenario may be a helpful explanation for the distribution of exoplanets (the dominance of super-Earths at $<1$ au \cite[]{Fressin:2013,Weiss:2014} and a possible peak in the occurrence of gas giants at $\sim2$--$3$ au \cite[]{Johnson:2010,Fernandes:2019}), as well as the architecture of the Solar System: rocky planets in the inner regions, gas giants in the middle, and icy giants in the outer regions.

The reduction of the pebble isolation mass in the inner region of the inviscid disk may cause the dichotomy between the inner super-Earths and outer gas giants \cite[]{Fung:2018}. However, the small pebbles (${\rm St}\lesssim10^{-3}$) may not be trapped at the local pressure maxima, and they continue to contribute to the growth of the planet \cite[]{Bitsch:2018}. The planet-induced flow isolation has the potential to explain the dichotomy even if the inward drift of small pebbles does not stop at the pressure maxima.
\section{Conclusions} \label{sec:conclusion}
Following \citetalias{Kuwahara:2020}, we investigated the influence of the 3D planet-induced gas flow on pebble accretion. We considered nonisothermal, inviscid gas flow and performed a series of 3D hydrodynamical simulations on a spherical polar grid that has a planet placed at its center. We then numerically integrated the equation of the motion of pebbles in 3D using hydrodynamical simulation data. Three-types of orbital calculation of pebbles are conducted in this study: in the unperturbed flow (\texttt{UP} case), in the planet-induced gas flow in the Stokes regime (\texttt{PI-Stokes} case), and in the planet-induced gas flow in the Epstein regime (\texttt{PI-Epstein} case). The subject of the range of dimensionless planetary mass in this study is $m=0.003$--$0.03$, which corresponds to a size ranging from three Moon masses to three Mars masses, $M_{\rm pl}=0.036$--0.36 $M_{\oplus}$, orbiting a solar-mass star at 1 au. Following \citetalias{Kuwahara:2020}, where only the shear regime of pebble accretion was considered, we extend our study to the headwind regime of pebble accretion in this study. Paper $\rm I\hspace{-.1em}I\,$ focused on the headwind regime of pebble accretion. We summarize our main findings as follows.

\begin{enumerate}
    \item The planet-induced gas flow can be divided into two regimes: the flow shear and flow headwind regimes. In the flow shear regime, the planet-induced gas flow has a vertically rotational symmetric structure \cite[]{Ormel:2015b,Fung:2015,Lambrechts:2017,Cimerman:2017,Kurokawa:2018,Kuwahara:2019,Bethune:2019,Fung:2019}. Gas from the disk enters the gravitational sphere at high latitudes and exits through the midplane region of the disk. The horseshoe flow lies across the orbit of the planet and has a columnar structure in the vertical direction. In the flow headwind regime, the headwind of the gas breaks the symmetric structure \cite[]{Ormel:2015b,Kurokawa:2018}. The planetary envelope is exposed to the headwind of the gas. Gas from the disk enters the gravitational sphere at low latitudes and exits at high latitudes of the gravitational sphere. The horseshoe flow lies inside the planetary orbit. We derived the flow transition mass analytically, $m_{\rm t,\,flow}$, which discriminates between the flow headwind and flow shear regimes.

    \item In the flow headwind regime, the trajectories of pebbles with ${\rm St}\lesssim10^{-3}$ in the planet-induced gas flow differ significantly from those in the unperturbed flow. The outflow in the fourth quadrant of the $x$-$y$ plane deflects the trajectories of pebbles. The pebbles are jammed outside the planetary orbit. The horseshoe flow shifts significantly to the negative direction in the $x$-axis. Because of the absence of the horseshoe flow and the outflow in the anterior region of the planetary orbit, pebbles passing near the planetary orbit are susceptible to becoming entangled in the recycling flow in the flow headwind regime. When pebbles enter the Bondi sphere and get entangled in the recycling flow, the relative velocity of pebbles is reduced by an order of magnitude because of the low speed of gas flow. This leads to an increase in the Bondi crossing time of pebbles. Thus, pebble accretion is enhanced in the planet-induced gas flow when pebbles are well coupled to the gas.
    
    \item From the relation between $m,\,m_{\rm t,\,flow}$ and $m_{\rm t,\,peb}$, we classify the results obtained in both \citetalias{Kuwahara:2020} and this study into four categories. When $m_{\rm t,\,peb}<m_{\rm t,\,peb}<m$ and $m_{\rm t,\,flow}<m<m_{\rm t,\,peb}$, we found that the outcome is identical to that in \citetalias{Kuwahara:2020}. When $m_{\rm t,\,peb}<m<m_{\rm t,\,flow}$, the influence of the planet-induced gas flow cannot be seen. In particular when $m<m_{\rm t,\,flow}<m_{\rm t,\,peb}$ and the Stokes gas drag law is adopted, pebble accretion is enhanced due to the planet-induced gas flow. The width of the accretion window, accretion cross section, and the accretion probability of pebbles in the planet-induced gas flow are larger than those in the unperturbed flow. 
\end{enumerate}
Following \citetalias{Kuwahara:2020}, we assumed that the global structure of the disk in terms of the distribution of the turbulence parameter and the size distribution of the solid materials in a disk based on previous studies \cite[]{Malygin:2017,Lyra:2019,Okuzumi:2019}. In Paper $\rm I\hspace{-.1em}I\,$, we considered an earlier phase of planet formation ($m\lesssim0.03$) compared to the what we considered in \citetalias{Kuwahara:2020}. We found that the planet-induced gas flow has little effect on pebble accretion for a range of the Stokes number assumed here ${\rm St}\geq10^{-3}$. Based on the discussion in \citetalias{Kuwahara:2020}, we conclude that the suppression of pebble accretion due to the planet-induced gas flow works in the late stage of planet formation ($m\gtrsim0.03$), in particular in the inner region of the disk ($\lesssim1$ au). We found that the dimensionless planetary masses of $0.03$ can be regarded as the planet-induced flow isolation mass, $m_{\rm PI,\,iso}$, in a specific case where ${\rm St}\lesssim10^{-3}$ and $\mathcal{M}_{\rm hw}\lesssim0.1$. This may be helpful to explain the distribution of exoplanets (the dominance of super-Earths at $<1$ au \cite[]{Fressin:2013,Weiss:2014} and a possible peak in the occurrence of gas giants at $\sim2$--$3$ au \cite[]{Johnson:2010,Fernandes:2019}), as well as the architecture of the Solar System, both of which have small inner, and large outer planets.

\begin{acknowledgements}
We thank Athena++ developers: James M. Stone, Kengo Tomida, and Christopher White. Numerical computations were in part carried out on Cray XC50 at the Center for Computational Astrophysics at the National Astronomical Observatory of Japan. A. K. was supported by JSPS KAKENHI Grant number 20J20681. H. K. was supported by JSPS KAKENHI Grant number 17H06457, 18K13602, 19H05072, and 19H01960.
\end{acknowledgements}



\newpage
\appendix
\setcounter{section}{1}
\def\thesection{A.\arabic{section}}
\setcounter{equation}{0}
\def\theequation{A.\arabic{equation}}
\setcounter{figure}{0}
\def\thefigure{A.\arabic{figure}}

\label{sec:appendix}
\subsection{Analytical estimation of pebble accretion}\label{sec:appendix1}
In the settling regime, the maximum impact parameter of accreted pebbles in the unperturbed flow is expressed by 
\begin{empheq}[left={b_{x}\simeq\empheqlbrace}]{alignat=2}
&b_{x,{\rm hw}}=\displaystyle2\sqrt{\frac{m{\rm St}}{\mathcal{M}_{\rm hw}}}, \quad &(m< m_{\rm t,\,peb}:  \text{headwind regime})\label{eq:b-hw}\\
&b_{x,{\rm sh}}=\displaystyle2\,{\rm St}^{1/3}\ R_{\rm Hill},  \quad &(m> m_{\rm t,\,peb}:\text{shear regime})\label{eq:b-sh}
\end{empheq}
\cite[]{Ormel:2010,Lambrechts:2012,Guillot:2014,Ida:2016,Sato:2016}, where $m_{\rm t,\,peb}$ is the transition mass for pebble accretion
\begin{align}
m_{\rm t,\,peb}=\frac{\mathcal{M}^{3}_{\rm hw}}{9{\rm St}}\label{eq:dim-less-transition}
\end{align}
\cite[]{Lambrechts:2012,Johansen:2017,Ormel:2017-pebble}. The dimensional pebble transition mass in the MMSN model can be described by
\begin{align}
M_{\rm t,peb}=1.67\times10^{-4}\,{\rm St}^{-1}\left(\frac{a}{1\,{\rm au}}\right)^{3/2}\,M_{\oplus}.\label{eq:pebble-transition}
\end{align}
In this study, we referred to the transition mass which divide the accretion regime of pebbles as "pebble transition mass" to distinguish pebble transition mass from flow transition mass, which determines the regime of the planet-induced gas flow (Eqs. (\ref{eq:m-flow-1}) and (\ref{eq:m-flow-2})).

\subsection{Additional figures}
\iffigure
 \begin{figure*}[htbp]
 \resizebox{\hsize}{!}
 {\includegraphics{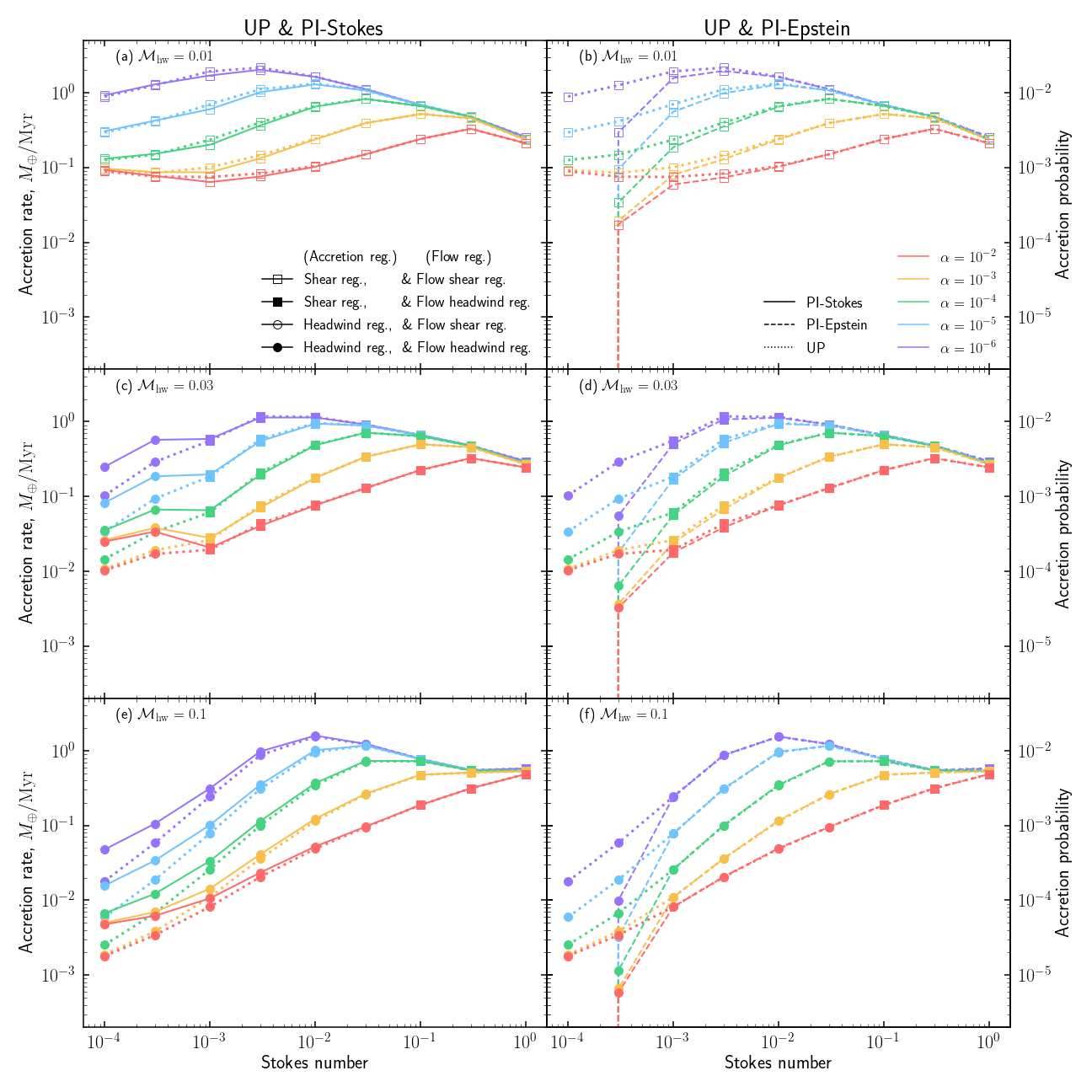}} 
 \caption{Same as \Figref{fig:Mdot-3D-m001}, but results obtained from \texttt{UP-m0003} case (dotted lines), \texttt{PI-Stokes-m0003} case (solid lines), and \texttt{PI-Epstein-m0003} case (dashed lines).}
\label{fig:Mdot-3D-m0003}
\end{figure*}
\fi
\iffigure
 \begin{figure*}[htbp]
 \resizebox{\hsize}{!}
 {\includegraphics{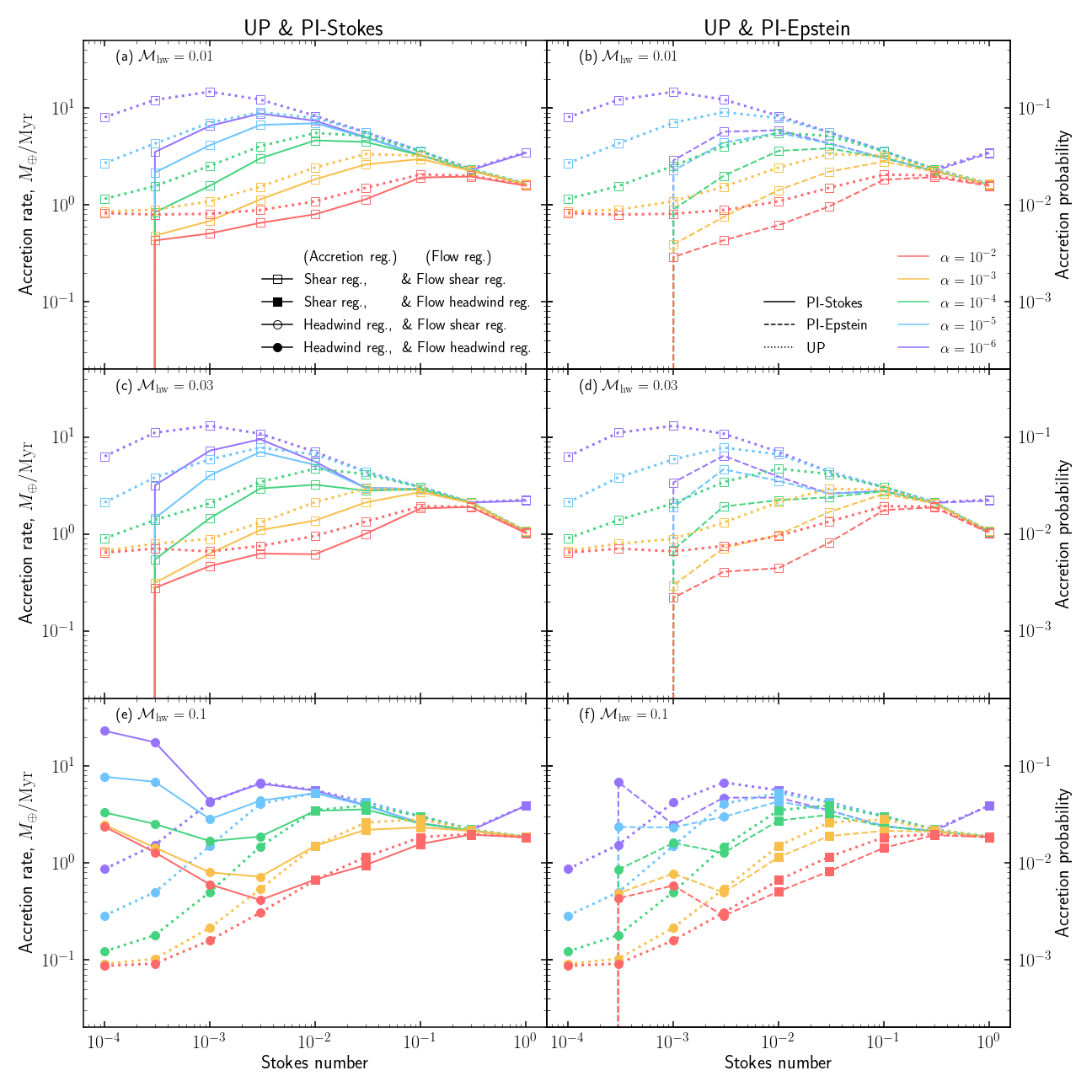}} 
 \caption{Same as \Figref{fig:Mdot-3D-m001}, but results obtained from \texttt{UP-m003} case (dotted lines), \texttt{PI-Stokes-m003} case (solid lines), and \texttt{PI-Epstein-m003} case (dashed lines).}
\label{fig:Mdot-3D-m003}
\end{figure*}
\fi
\end{document}